\documentclass[aip,jcp,twocolumn,10pt]{revtex4-1}
\usepackage{times}
\usepackage{amsmath}
\usepackage{amsfonts}
\usepackage{graphics,graphicx}

\usepackage{color}
\usepackage{tabularx}
\usepackage[T1]{fontenc}

\newcommand{\degree}{$^\circ$}

\newcommand{\citen}[1]{\onlinecite{#1}}

\begin{document}  

\title{Tuning Chemical Precompression: Theoretical Design and Crystal Chemistry of Novel Hydrides in the Quest for Warm and Light Superconductivity at Ambient Pressures}

\author{Katerina P. Hilleke}
\affiliation{Department of Chemistry, State University of New York at Buffalo, Buffalo, NY 14260-3000, USA}
\author{Eva Zurek}\email{ezurek@buffalo.edu}
\affiliation{Department of Chemistry, State University of New York at Buffalo, Buffalo, NY 14260-3000, USA}

\begin{abstract}

Over the past decade, a combination of crystal structure prediction techniques and experimental synthetic work has thoroughly explored the phase diagrams of binary hydrides under pressure. The fruitfulness of this dual approach is demonstrated in the recent identification of several superconducting hydrides with $T_c$s approaching room temperature. We start with an overview of the computational procedures for predicting stable structures and estimating their propensity for superconductivity. A survey of phases with high $T_c$ reveals some common structural features that appear conducive to the strong coupling of the electronic structure with atomic vibrations that leads to superconductivity. We discuss the stability and superconducting properties of phases containing two of these -- molecular H$_2$ units mixed with atomic H and hydrogenic clathrate-like cages -- as well as more unique motifs. Finally, we argue that ternary hydride phases, which are far less-explored, are a promising route to achieving simultaneously superconductivity at high temperatures and stability at low pressures. Several ternary hydrides arise from the addition of a third element to a known binary hydride structure through site mixing or onto a new site -- and several more are based on altogether new structural motifs.

\end{abstract}

\maketitle

\section{Introduction}

Many diamonds have been broken, and CPU hours have been burned. Yet, at long last the elusive goal of high-temperature superconductivity in light element systems, as originally proposed by Ashcroft in 1968 \cite{Ashcroft:1968a}, is at hand. Record-breaking superconducting  critical temperatures ($T_c$s) have been measured in a handful of exotic hydrides such as H$_3$S (203~K at 155~GPa) \cite{Drozdov:2015a}, LaH$_{10}$ (250-260~K near 200~GPa) \cite{Zulu:2018-La,Drozdov:2019}, YH$_9$ (240-260~K at 180-200~GPa \cite{Snider:2021, Kong:2021a}), YH$_6$ ($\sim$220~K at 165-185~GPa \cite{Troyan:2021a, Kong:2021a}), and most notably a $T_c$ of 288~K at 267~GPa has been reported in a C-S-H phase \cite{Snider:2020}.  These materials defy classical chemical structure and bonding concepts, and challenge the notion that high $T_c$s are the domain of unconventional superconductors. In many cases, their discovery was theoretically inspired, contrary to Matthias' sixth rule for guiding a successful search for new superconducting materials, ``Stay away from theorists!'' \cite{Conder:2016}. 

\begin{figure}
\begin{center}
\includegraphics[width=1\columnwidth]{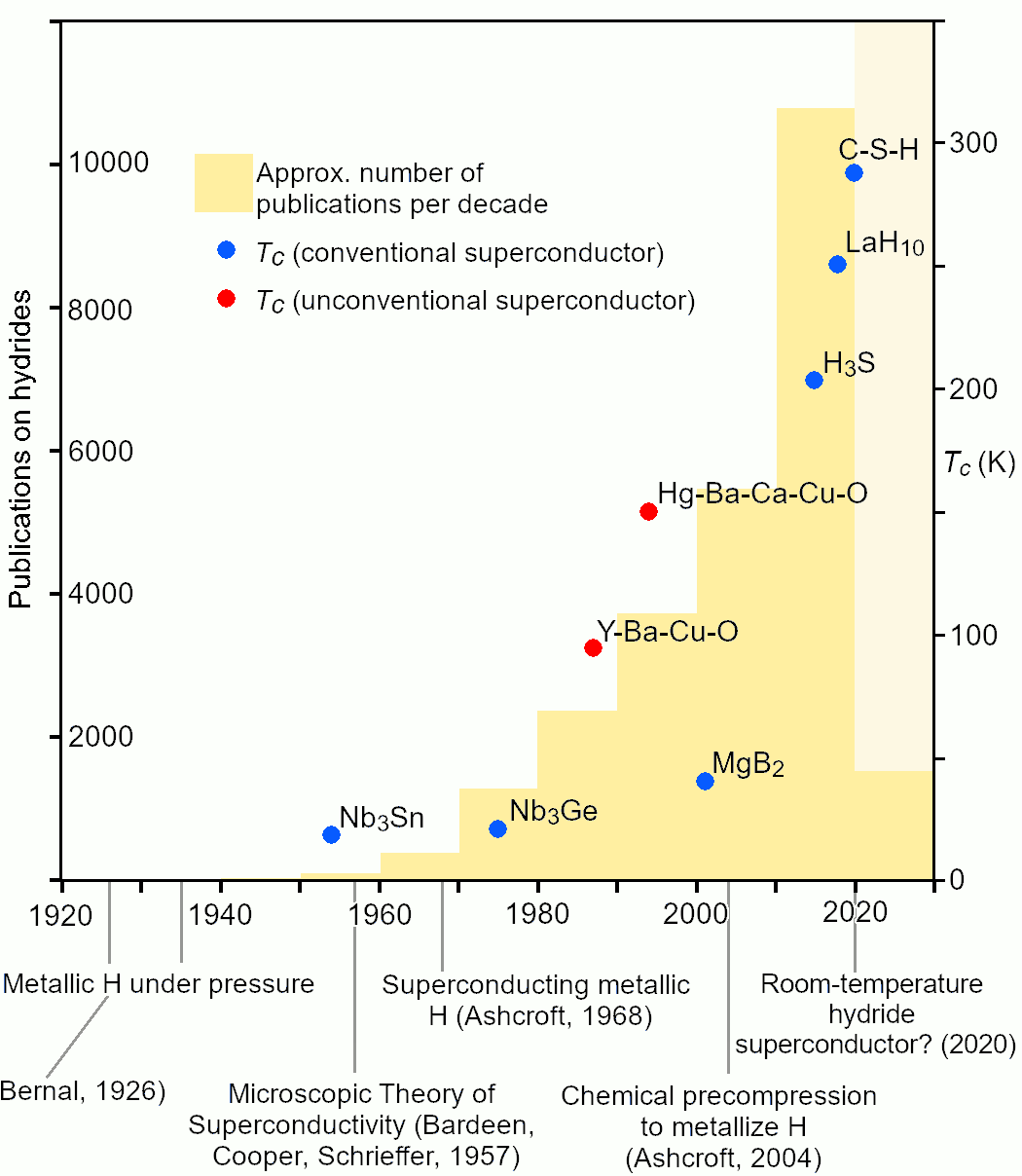}
\end{center}
\caption{Progress and interest in the field of superconducting hydrides over time. The approximate number of publications involving hydride superconductivity (left \emph{y}-axis, yellow bars) per decade has grown rapidly, with landmark publications highlighted below. Key experimentally measured $T_c$ values (right \emph{y}-axis, blue dots for conventional  superconductors and red dots for unconventional superconductors) increase from modest temperatures in Nb$_3$Sn to near room-temperature superconductivity in the high pressure LaH$_{10}$ and C-S-H phases.}
\label{fig:hydride_rush}
\end{figure}

The obsession with squeezing hydrogen until it metallizes can be traced back to a 1926 proposal by J.D.\ Bernal. This idea was communicated in the literature for the first time by Wigner and Huntington, who proposed that at  25~GPa hydrogen, the first element of the periodic table, would follow its neighbors in group I to adopt a metallic solid structure \cite{Wigner:1935,note}. Ashcroft argued that in such a state the light mass of hydrogen, with its concomitant high phonon frequencies, large concentration of electronic states near the Fermi level, and strong electron phonon coupling arising from the lack of screening by core electrons would yield  a superconducting phase that could maintain its character up to very high temperatures \cite{Ashcroft:1968a}. At the time, the highest known $T_c$, of 18~K, was exhibited by the A15 intermetallic Nb$_3$Sn~\cite{Matthias:1954}, and it was overtaken in 1975 by the isotypic Nb$_3$Ge with $T_c$ up to 22~K in sputtered thin films~\cite{Gavaler:1973} (Figure \ref{fig:hydride_rush}). Far from achieving superconductivity at room temperature, these $T_c$s did not even approach the boiling point of nitrogen (77~K). Thus, the race to metallize hydrogen and harness its remarkable putative superconductivity was on. 

For a while, it seemed as though a completely different family of materials, the superconducting cuprates, could be the first to break the room-temperature barrier, with a Y-Ba-Cu-O phase that possessed a $T_c$ of 93~K being the first to surpass the boiling point of liquid nitrogen \cite{Wu:1987}. Unlike Nb$_3$Sn, Nb$_3$Ge, and the recently synthesized metal hydride superconductors, this family of materials belongs to a class of unconventional superconductors whose superconducting mechanism, though not fully understood, cannot be explained by Bardeen-Cooper-Schrieffer (BCS) theory.  

Despite the continuing increase of the predicted pressure of hydrogen metallization~\cite{Narayana:1998a,Loubeyre:2002,Zha:2013a,Eremets:2011a,Howie:2012a,Simpson:2016a}, keeping elemental hydrogenic superconductivity just out of reach, a new strategy to lower the metallization pressure was proposed. Ashcroft suggested hydrogen dominant alloys as plausible candidates for achieving metallicity -- and superconductivity -- at lower pressures than elemental hydrogen, owing to the \emph{``chemical precompression''} from the alloying element.~\cite{Ashcroft:2004a} The group 14 hydrides such as SiH$_4$ and GeH$_4$ were first suggested, in part because they possessed four completely occupied bands, similar to the recently discovered light-element conventional superconductor MgB$_2$~\cite{Nagamatsu:2001}, as well as the propensity of the tetrels towards superconductivity (albeit under pressure, in the case of Si and Ge).

A plethora of studies on the Si-H \cite{Feng:2006a,Pickard:2006a,Yao:2007-Si,Kurzydlowski:2011a} and Ge-H \cite{Canales:2006a,Gao:2008a} systems were promptly undertaken -- including in the nascent field of \emph{ab initio} crystal structure prediction (CSP), which has been shown to be useful in identifying a variety of stable and metastable structures. At elevated pressures where the chemical intuition of the ambient-pressure-trained minds of most scientists starts to fail \cite{Zurek:2019k,Grochala:2007a}, computational CSP techniques were especially helpful \cite{Zurek:2014i}. Not only are stoichiometries unimaginable at ambient pressure predicted to become stable, including NaCl$_7$~\cite{Zhang:2013} and H$_4$I~\cite{Zurek:2015d}, but strange new structures comprised of curious motifs such as fused H$_5$ pentagons or clathrate-like cages composed of hydrogen can appear \cite{Zurek:2018m}. 

Ultimately, this line of research resulted in the discovery of the hydride phases with record-breaking $T_c$s mentioned in our opening.  But, much more work remains to be done! The remarkable superconducting properties emerge only at pressures approaching those experienced at the center of the Earth. Few high-$T_c$ hydrides are predicted to remain stable at pressures below one megabar, and over the past decade most high-pressure binary hydride phase diagrams have already been exhaustively investigated. \cite{Zurek:2018d,Flores-Livas:2020} The same cannot be said for ternary phase diagrams. What's more, it has been suggested that increasing complexity could result in phases that possess both high $T_c$s and the ability to be quenched to lower pressures. The latter property represents an extension of the chemical precompression idea: in the same way that a metal atom could ``chemically precompress'' pure hydrogen, stuffing a third element into a binary hydride might render it stable to lower pressures. The staggeringly vast phase space of ternary hydride materials is only beginning to be systematically explored, and it has been proposed that chemical concepts and machine-learning approaches will be useful in guiding such searches~\cite{Hutcheon:2020,2021roadmap,Shipley:2021}. 

In this Perspective, we review superconductivity in hydrides stable at elevated pressures, with particular focus on the alkaline and rare earth metal binary hydride phases as well as on ternary hydride systems that can be constructed by alloying these binaries with another metal or a main group element.  In section~\ref{sec:comp} we provide an overview of the computational techniques and methodologies employed in current computational studies of high-pressure hydrides, first covering \emph{ab initio} crystal structure prediction, then the equations and procedures used to estimate $T_c$s. In section~\ref{sec:motifs} we present the  binary phases, grouping them into families based on structural motifs of the hydrogen sublattices -- mixing of molecular and atomic hydrogen, clathrate-like cages, and additional curious motifs including fused H$_5$ pentagons. Ternary hydride materials ranging from substitutions of a third element into a binary hydride phase to altogether new structures are covered in Section~\ref{sec:ternaries}, and we close with a brief outlook in Section~\ref{sec:outlook}.

\section{Computational Techniques} \label{sec:comp}

\subsection{Computational Methodologies} \label{sec:methods}

At the pressures required to stabilize many of the superconducting polyhydrides, the chemical intuition developed from living at atmospheric pressure breaks down. Seemingly bizarre stoichiometries such as NaCl$_3$~\cite{Zhang:2013} and FeH$_{5}$~\cite{Pepin:2017a} have been synthesized in diamond anvil cell experiments.  Meanwhile, geometrical motifs that are not observed at 1~atm such as the pentagraphene units in ScH$_{10}$ \cite{Zurek:2019k} and HfH$_{10}$~\cite{Xie:2020}, as well as the hydrogenic clathrate cages in CaH$_{6}$~\cite{Wang:2012} and LaH$_{10}$~\cite{Peng:Sc-2017,Liu:2017-La-Y} have been predicted. Pressure also influences the electronic structure and physical properties of matter. For example, in a behavior reversal, whereas hydrogen is expected to become metallic and superconducting,~\cite{Wigner:1935,Ashcroft:1968a,Mcmahon:2011b} sodium undergoes a metal-to-insulator transition under pressure. \cite{Ma:2009a}

These are some of the reasons \emph{a priori} CSP techniques have emerged as the methods of choice to identify promising synthetic targets computationally.~\cite{Wang:2014a,random3,Zurek:2014i,Zurek:2014d} What's more, high pressure experiments may be difficult to execute, the products made can be challenging to characterize, the material properties hard to interrogate, and the results subject to interpretation. As computational techniques improve, CSP may become increasingly important in identifying synthetic targets for useful properties ranging from superhardness~\cite{Zurek:2019b} to superconductivity~\cite{Zurek:2016j}. Moreover, CSP is steadily gaining importance in helping to characterize already synthesized materials whose structures are unknown. 

Several techniques for predicting global and important local minima of a crystalline material have been leveraged to great success. The global optimization schemes used towards this end are stochastic in nature, with no guarantee of finding the true global minimum within a potential energy surface (PES) that can be highly complex and contain several competitive local minima. The techniques that have been developed are thus based on several metaheuristics designed to find good solutions for optimization problems. These include minima hopping~\cite{Goedecker:2004,Amsler:2010,Amsler:2019}, simulated annealing~\cite{Kirkpatrick:1983}, metadynamics~\cite{Laio:2002,Sun:2009}, particle swarm optimization~\cite{Call:2007,Wang:2015}, genetic (evolutionary) algorithms~\cite{uspex5,Zurek:2011a}, and (quasi-) random searches~\cite{Pickard:2009,random2}. Whereas the first three are best suited for local exploration, the latter three can broadly traverse the PES. Quantifying the performance of an algorithm is difficult due to their stochastic nature, however it has been shown that various options can greatly enhance the success rate \cite{Zurek:2020i}.

Over the course of a CSP search, a single run can generate hundreds to thousands of structures. Often, multiple runs are required to ensure the global minimum has been found -- and furthermore, CSP is often performed at various pressures. All of this is compounded by the fact that identification of stable phases requires generation of a full convex hull of a system. While some CSP techniques construct the convex hull on the fly, others require individual searches be performed for all chosen stoichiometries. Each of these thousands of structures is then optimized, typically using density functional theory (DFT), to the nearest stationary point. With the exception of dynamics based schemes, CSP explores the 0~K PES, as calculating the vibrational contributions to the free energy requires a significant increase in computational power relative to local relaxation. 

Geometry optimizations are usually performed within DFT using the generalized gradient approximation (GGA) functional family, commonly that of Perdew, Burke, and Ernzerhof~\cite{Perdew:1996a}. Recently, some CSP techniques have employed machine learned interatomic potentials that can be parameterized via active learning, allowing for local optimizations to take place on greatly reduced timescales relative to DFT~\cite{adaptive:2014a,Tong:2018,Podryabinkin:2019,Hong:2020}, however, these are not yet widely implemented.  On the other hand, complex electronic structures may necessitate the use of \emph{meta}-GGA functionals such as SCAN \cite{Zurek:2021a} or treatment of dispersion~\cite{Kaewmaraya:2011a,Deringer:2014,Roman-Roman:2015}. However, functionals that go beyond GGA are typically employed only to get a more accurate energy ordering after the CSP search is complete~\cite{Hermann:2016,Semenok:2021,Zhao:2020,Hai:2021}, to manage computational expense. Since GGA-type functionals tend to underestimate metallization that results from pressure induced band broadening, hybrid functionals such as HSE06~\cite{Krukau:2006} are often employed to verify metallicity.

Structures of a similar composition are ranked by enthalpy (for high pressure calculations) or internal energy (for ambient pressure calculations). The relevant thermodynamic quantity is plotted versus composition to trace out the convex hull. The phases that lie on the hull are thermodynamically stable, but dynamic stability must also be checked by calculating phonon dispersions and confirming the absence of imaginary phonon modes. At this point, additional corrective factors to the hull can be considered, including the zero-point energy (ZPE) and the role of anharmonic effects. The latter is quite onerous to calculate, so it is typically neglected. However, it has been illustrated that anharmonic effects can alter the PES and superconducting properties of many of the high temperature superconducting hydrides~\cite{Errea:2015a,Errea:2020,Hou:2021}. Meanwhile, the light mass and resultant high frequency atomic vibrations associated with hydrogen tend to culminate in substantial ZPE contributions. As a consequence, the inclusion of ZPE can alter the pressure range within which a given phase is stable, or stabilize one structure over another. Thus, ZPE effects are typically included in final determinations of a convex hull or pressure-stability ranges of different structures. 

Let us now summarize the computational workflow for the discovery of superconducting hydrides. After choosing a system for study, CSP runs at various pressures are carried out to find the most stable configurations as a function of composition. A number of the top candidates are chosen for a second round of local optimizations with higher accuracy settings, obtaining finalized lattices and enthalpies of formation. Calculations of phonon dispersions show which phases are dynamically stable and allow for the inclusion of ZPE into the 0~K enthalpy of formation. A convex hull is plotted -- often with and without the ZPE -- indicating which structures are expected to be stable (or low-lying metastable) at various pressures. Metastable phases are often reported because the real experimental conditions used for the synthesis can stabilize them. The following may guide the selection of structures to keep: a datamining analysis of the Materials Project database revealed that the median metastability (by DFT calculation) of all known inorganic crystalline materials was roughly 15 meV/atom, going up to roughly 70 meV/atom at the 90$^{th}$ percentile.~\cite{Sun:2016} The electronic structure of these phases is examined to determine which ones may have a large propensity for superconductivity (those with a high density of states at the Fermi level, DOS at $E_F$), and estimates of their $T_c$, using the techniques described in the next secion, are made.

\subsection{Calculating the Critical Temperature} \label{sec:Tc}
In 1911, Onnes detected an abrupt disappearance in the resistivity of Hg metal at very low temperatures~\cite{Onnes:1911}, sparking investigations into the cause of this phenomenon, termed superconductivity. 
An important hint came from the discovery of the isotope effect~\cite{Maxwell:1950,Reynolds:1950,Frohlich:1950}, where the onset of superconductivity could be altered merely by the atomic mass of an element. 
This suggested that the vibrations of a crystalline lattice, which are dependent on atomic mass, may be important for the superconducting mechanism. In 1957, Bardeen, Cooper, and Schrieffer (BCS) published their \emph{Microscopic Theory of Superconductivity}~\cite{bcs1,bcs2}, describing the underlying theory for what are now known as conventional superconductors. 
These materials include intermetallic compounds such as Nb$_{3}$Sn~\cite{Matthias:1954} and MgB$_{2}$~\cite{Nagamatsu:2001}, as well as the hydrogen-based materials we are concerned with here, such as  H$_{3}$S~\cite{Duan:2014,Drozdov:2015a} and LaH$_{10}$~\cite{Peng:Sc-2017,Liu:2017-La-Y,Zulu:2018-La,Drozdov:2018-La}. 

BCS theory postulates that the coupling of the motion of an electron with a phonon can overcome the Coulomb repulsion between the two negatively charged particles, resulting in an electron pair that is a composite boson. Typically, this interaction is illustrated in the following scenario. As an electron moves within a crystalline lattice, the Coulombic attraction between the electron and the positive ion cores leads to a momentary distortion of the lattice towards the electron, resulting locally in a slightly higher positive charge. A second electron, with opposite spin and momentum to the first, is attracted to this region. At great distances, this attractive force can overcome the Coulombic repulsion between a pair of negatively charged electrons, resulting in a paired state termed a Cooper pair. Although net attractive, this interaction between two electrons remains quite weak and can be broken with sufficient thermal energy. The critical temperature, $T_{c}$, marks the temperature above which the Cooper pairs are destroyed and the phase is no longer superconducting. 

Within BCS theory $T_{c}$ can be obtained via:
\begin{equation}
 k_BT_{c} = 1.14\hbar\omega\exp\bigg[{\frac{-1}{N(0)V}}\bigg] 
\label{eq:tc1} \end{equation}
where $\omega$ is the average phonon frequency of the system, $N(0)$ is the single spin DOS at $E_F$, and  $V$ is the pairing potential from the electron-phonon interaction. This expression, however, does not account for the damping effect in the electron-phonon interaction that arises because the displacements of the heavier ions associated with the phonon modes are slower than the motion of the light electrons. 

To account for this, the Eliashberg equations~\cite{eliashberg:1960}, based on a Green’s function formalism, include the damping effect and can be solved numerically. A key quantity, termed the Eliashberg spectral function, $\alpha^{2}F(\omega)$, can be computed via
\begin{equation}
 \alpha^{2}F(\omega) = \frac{1}{2\pi N(0)} \sum_{qj}\frac{\gamma_{qj}}{\omega_{qj}} \delta(\hbar\omega - \hbar\omega_{qj}) ,
\label{eq:a2F} \end{equation}
with phonon linewidth, $\gamma_{qj}$, and frequency, $\omega_{qj}$ for a phonon $j$ with a wavevector $q$. 

McMillan performed numerical solution of the Eliashberg equations for 22 systems to determine a simple expression that could be used to estimate $T_c$~\cite{McMillan:1968}. The phonon DOS of Nb was used to model the shape of the Eliashberg spectral function (Equation~\ref{eq:a2F}), yielding the simplified formula
\begin{equation}
  T{_c} = \frac{\Theta_D}{1.45} \text{exp}\bigg[-\frac{1.04(1+\lambda)}{\lambda-\mu^*(1+0.62\lambda)}\bigg] 
\label{eq:AD} \end{equation} 
based on the Debye frequency, $\Theta_{D}$, and the renormalized Coulomb repulsion parameter, $\mu^{*}$. The electron-phonon coupling constant, $\lambda$, is derived from the Eliashberg spectral function according to 
\begin{equation}
  \lambda = 2 \int_0^{\infty} \frac{\alpha^2F(\omega)}{\omega}\text{d}\omega .
\label{eq:lambda} \end{equation}

In cases where the phonon spectrum of the system differed greatly from that of Nb -- in Hg, for example -- the electron phonon coupling (EPC) constant $\lambda$ extracted from tunneling experiments differed from the value of $\lambda$ obtained by solving Equation~\ref{eq:AD} given experimental values of $T_{c}$ and $\Theta_{D}$. Thus, for systems with $\lambda$ > 1 the McMillan equation often failed to accurately capture the superconducting behavior.

Subsequently, Allen and Dynes published a modified version of the McMillan equation~\cite{Allen:1975} to estimate $T_{c}$:
\begin{equation}
  T{_c} = \frac{\omega_\text{ln}}{1.2} \text{exp}\bigg[-\frac{1.04(1+\lambda)}{\lambda-\mu^*(1+0.62\lambda)}\bigg] .
\label{eq:ADM} \end{equation} 
This equation was based on 217 numerical solutions of the Eliashberg equations for $0 \leq \mu^{*} < 0.2$, $0.3 < \lambda < 10$, and $\alpha^{2}F(\omega)$ shapes of Pb and Hg obtained from the known tunneling experiments, as well as an Einstein model, which includes no Coulombic repulsion ($\mu^*$=0) and whose shape of $\alpha^{2}F(\omega)$ maximizes $T_c$ for given values of $\lambda$ and $\omega_\text{ln}$.

In an earlier work, Dynes had replaced the $\frac{\Theta_D}{1.45}$ prefactor in the McMillan equation (Equation \ref{eq:AD}) with a prefactor of $\frac{<\omega>}{1.2}$.~\cite{Dynes:1972} In the Allen-Dynes modified McMillan equation (Equation \ref{eq:ADM}) <$\omega$> was further replaced with the logarithmic average phonon frequency, $\omega_\text{ln}$, calculated as 
\begin{equation}
  \omega_\text{ln} = \text{exp}\bigg[\frac{2}{\lambda}\int_0^{\infty}\frac{d\omega}{\omega}\alpha^{2}F(\omega)\text{ln}(\omega)\bigg] .
\label{eq:omegalog} \end{equation}
This procedure was found to drastically reduce the dependence of $T_c$ on the shape of the spectral function, resulting in an expression that is quite accurate for predicting $T_c$ for materials with $\lambda<1.5$. 

In a computational study, the Eliashberg spectral function can be calculated for a material using DFT codes including the Quantum Espresso~\cite{Giannozzi_2009}, ABINIT~\cite{Gonze2020}, and EPW~\cite{Ponce:2016} packages from which values for $\lambda$ can be obtained and $T_{c}$ can be estimated via Equation~\ref{eq:ADM}. Typically, a range of $T_{c}$ values corresponding to solutions where $0.1 \leq \mu^{*} \leq 0.13 \text{ or } 0.15$ is given in an article. In strongly coupled systems, however, where $\lambda$>1.5, the $T_{c}$ estimated with Equation~\ref{eq:ADM} is too low. In these cases, the usual course is either numerical solution of the Eliashberg equations~\cite{eliashberg:1960} to generate a more appropriate estimate of $T_{c}$, or the application of correction factors $f_1$ (for strong coupling) and $f_2$ (for shape dependence).~\cite{Allen:1975} The approximate form for these factors is given by
\begin{equation}
  f_1 = \bigg\{ 1+ \bigg[ \frac{\lambda}{2.46(1+3.8\mu^*)}  \bigg]^{\frac{3}{2}} \bigg\}^{\frac{1}{3}},
\label{eq:f1} \end{equation}
and
\begin{equation}
  f_2 =  1+ \frac{(\omega_2/\omega_\text{ln}-1)\lambda^2)}{\lambda^2+[1.82(1+6.3\mu^8)(\omega_2/\omega_\text{ln})]^2}   
\label{eq:f2} \end{equation}
where the logarithmic average of the phonon frequency $\omega_\text{ln}$ is calculated as in Equation~\ref{eq:omegalog} and $\omega_2$ is

\begin{equation}
  \omega_2 =  \bigg[\frac{2}{\lambda}\int_0^{\infty}\omega\alpha^2F(\omega)\text{d}\omega \bigg]^{1/2}   .
\label{eq:w2} \end{equation}

An alternative way to calculate $T_c$ is via density functional theory for superconductors (SCDFT)~\cite{Oliveira:1988}. By creating a density functional framework based on  the system electron density, N-body nuclear interactions, and the superconducting order parameter~\cite{Oliveira:1988,Luders:2005a,Marques:2005a}, it does not rely on empirical parameters such as the effective Coulombic repulsion term $\mu^*$.

\section{Hydrogenic Motifs} \label{sec:motifs}

\subsection{Mixed Molecular and Atomic Hydrogen} \label{sec:mixed}

\begin{figure}
\begin{center}
\includegraphics[width=1\columnwidth]{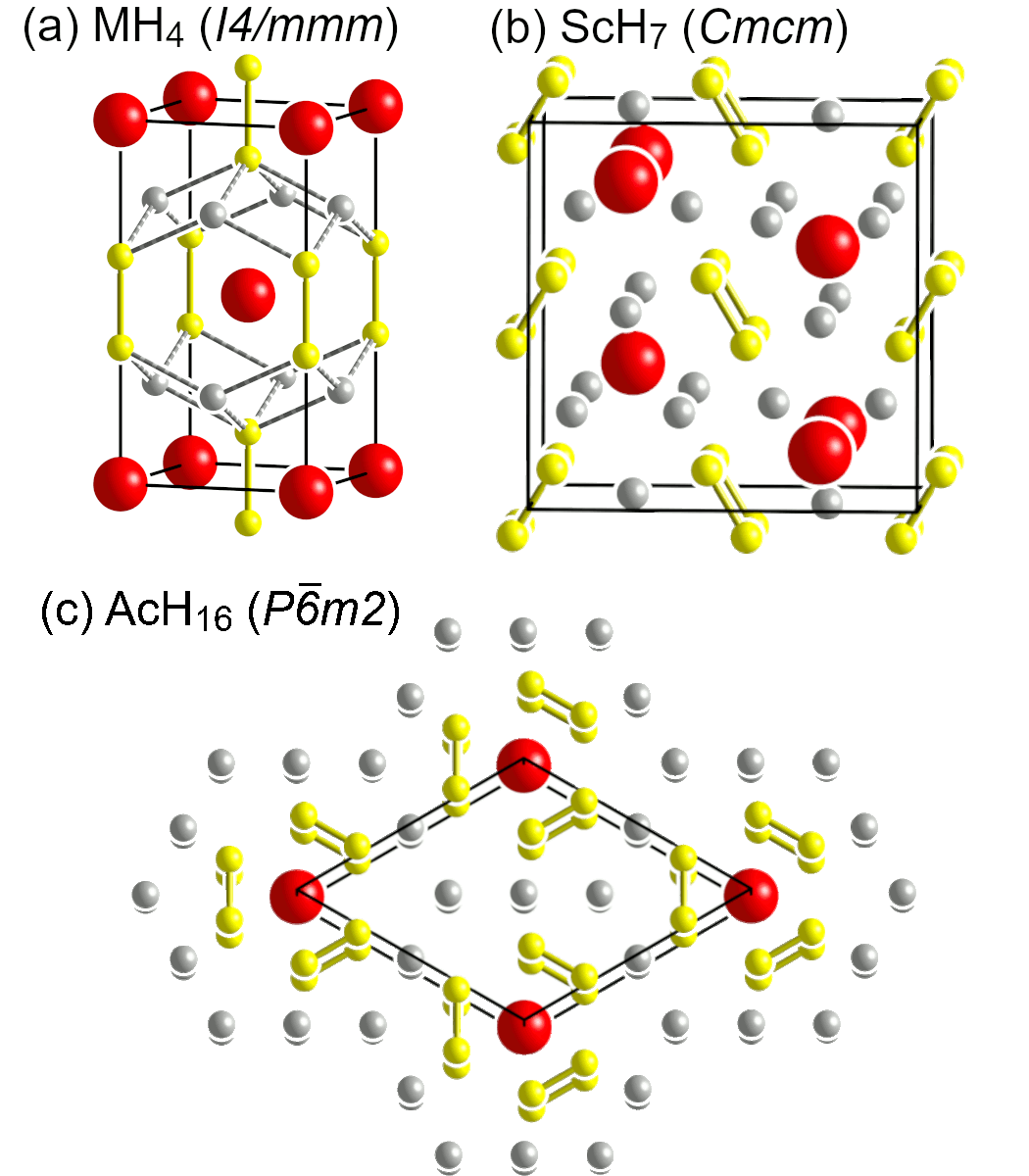}
\end{center}
\caption{Structures of metal hydride phases with atomic H and molecular H$_2$ units with metal atoms in red, atomic H in grey, and molecular H$_2$ units in yellow. (a) $I4/mmm$ structure common to several MH$_4$ phases has apical hydrogens, H$_a$, in molecular units and atomic basal hydrogens, H$_b$. (b) ScH$_7$ structure ($Cmcm$) with alternating layers of H$_2$ units and atomic H with Sc. (c) $P\bar{6}m2$ AcH$_{16}$ structure with three H$_2$ units above and below each Ac, which is also coordinated to 12 atomic H.}
\label{fig:mixed_struct}
\end{figure}

Within these hydride phases, the hydrogen sublattice can assume many forms -- atomic/hydridic hydrogen, molecular H$_2$ and H$_3^-$ (and more exotic) units, weakly covalent clathrate cages, and more. One ubiquitous group of phases are those with a mixture of atomic H and molecular H$_{2}$. Among these, the presence of distinct hydrogenic motifs is associated with varying degrees of electron transfer to and from the molecular and atomic units, which strongly influences the electronic structure and superconducting properties of the resulting phases. An immensely common structure type among metal hydride phases is the $I4/mmm$ symmetry phase (Figure~\ref{fig:mixed_struct}a) of which ThCr$_2$Si$_2$ and BaAl$_4$ are archetypes. In this geometry, electropositive metal atoms occupy the $2a$ Wyckoff positions and hydrogen atoms are distributed between the apical (Wyckoff position $4e$) and basal (Wyckoff position $4d$) sites. Within ThCr$_2$Si$_2$ the Si/Cr atoms lie on the $4e/4d$ sites.

Depending on the volume of the unit cell and amount of electron transfer between the metallic and hydrogenic lattices, the apical hydrogens, H$_a$, can be treated as weakly bonded to nearly hydridic in character, while the basal hydrogens, H$_b$, are generally formally considered hydridic, so that the overall distribution of charge can be expressed as $\text{M}^{x+}(\text{H}_b^{-})_2(\text{H}_a)_2^{(x-2)-}$. This phase has been predicted to lie on the convex hull of numerous metal-hydrogen binary systems: CaH$_4$~\cite{Wang:2012,Zurek:2018c,Shao:2019}, SrH$_4$~\cite{Hooper:2013,Wang:2015a}, ScH$_4$~\cite{Abe:Sc-2017,Qian:Sc-2017,Zurek:2018b}, YH$_4$\cite{Li:2015a,Liu:2017-La-Y}, LaH$_4$\cite{Liu:2017-La-Y}, CeH$_4$\cite{Peng:Sc-2017,Li:2019,Salke:2019}, PrH$_4$~\cite{Peng:Sc-2017,Zhou:2020b,PenaAlvarez:2020}, PuH$_4$~\cite{Zhao:2020} EuH$_4$~\cite{Semenok:2020}, NdH$_4$~\cite{Zhou:2020,Peng:Sc-2017}, TbH$_4$~\cite{Hai:2021}, ThH$_4$~\cite{Kvashnin:2018,Semenok:2020} and ZrH$_4$\cite{Abe:2018-Zr}. In addition, model convex hulls based on computed enthalpies for a series of ubiquitous structure types observed in metal hydride structures found PmH$_4$, SmH$_4$, EuH$_4$, GdH$_4$, TbH$_4$, DyH$_4$, HoH$_4$, ErH$_4$, TmH$_4$, and LuH$_4$ to be stable~\cite{Peng:Sc-2017}.

The staggering array of binary systems for which this structure is stable lends itself well to probing property trends based on character of the metal element, as seen in a recent overview.~\cite{Bi:2021} Oxidation states for M$^{x+}$ range from 2+ to 4+, with CaH$_4$ and ThH$_4$ as exemplars, resulting in different amounts of electron density donated to the hydrogen lattice. The added electron density weakens the H$_a$-H$_a$ bonds via a two-pronged mechanism involving population of the H$_2$ $\sigma_u^*$ antibonding orbitals as well as a Kubas-like interaction of coupled H$_2$ $\sigma_g~\rightarrow$ M $d$ and M $d~\rightarrow$ H$_2$ $\sigma_u^*$ donation. Thus, the H$_a$-H$_a$ bond distance depends greatly on the valency of the metal atom, with the divalent tetrahydrides having H$_a$-H$_a$ distances slightly longer but comparable to molecular H$_2$, while the tetravalent hydrides have H$_a$-H$_a$ distances nearly twice as long (with four electrons per formula unit donated to the hydrogen lattice, each hydrogen will have a formal charge of -1, corresponding to hydridic character). Hoffmann and Zheng have shown that the X-X bond in the AB$_2$X$_2$ structure type can be tuned by the choice of the transition metal atom~\cite{Hoffmann:1985,Hoffmann:1987a}, similar to what we find here for the H$_a$-H$_a$ contact.

The distance dependence of the H$_a$-H$_a$ contact on the valence and size of the metal atom has noteworthy consequences on the predicted $T_c$s, whose values are recorded in Table~\ref{tab:mixed}. In MgH$_4$, which has an estimated $T_c$ of 126~K at 300~GPa~\cite{Bi:2021}, a major contributor to the EPC constant $\lambda$ involved libration of the H$_a$-H$_a$ units coupled to interactions with the H$_b$ network. One phonon mode with an especially large EPC contribution involved circular motions of the H$_b$ atoms which, upon decreasing pressure, softens to the point of instability. Often, superconductivity is favored near the onset of dynamic instability.  In YH$_4$, where the metal is formally trivalent, strong contributions to the EPC arise along a soft phonon mode around the N point that is related to the motions of the H$_a$ atoms~\cite{Li:2015a}. Large contributions to the EPC within ScH$_4$ arise mostly from motions of the H$_a$ atoms, one of which involves a similar coupling of a H$_a$-H$_a$ libration to the H$_b$ network as in MgH$_4$~\cite{Bi:2021}.

\begin{table*}[!ht]
    \centering
    \def\arraystretch{1}
    \caption{Superconducting properties of select binary hydrides with mixed molecular and atomic hydrogen (Figure \ref{fig:mixed_struct}).}
     \setlength{\tabcolsep}{3mm}{        
       \begin{tabular}{c c c c c c}
\hline
Structure & Pressure (GPa) & $\lambda$ & $\mu^*$ & $T_c$ \\
(space group) & & & & \\
\hline
ScH$_{4}$~\cite{Bi:2021} ($I4/mmm$) & 120 & 1.19 & 0.13-0.1 & 113.3-101.5$^a$, 92.3-81.8$^b$\\
ScH$_{4}$~\cite{Zurek:2018b} ($I4/mmm$) & 120 & 1.68 & 0.1 & 163$^a$, 92$^b$ \\
ScH$_{4}$~\cite{Abe:Sc-2017} ($I4/mmm$) & 195 & 0.89 & 0.13-0.1 & 67-81$^b$ \\
ScH$_{4}$~\cite{Qian:Sc-2017} ($I4/mmm$) & 200 & 0.99 & 0.1 & 98$^b$ \\
ZrH$_{4}$~\cite{Bi:2021} ($I4/mmm$) & 150 & 1.24 & 0.13-0.1 & 77.1-66.1$^a$, 58.9-52.4$^b$\\
ZrH$_{4}$~\cite{Abe:2018-Zr} ($I4/mmm$) & 230 & 0.89 & 0.19 & 47$^a$ \\
MgH$_{4}$~\cite{Bi:2021} ($I4/mmm$) & 255 & 1.05 & 0.13-0.1 & 124.7-105.0$^a$, 112.9-98.4$^b$\\
MgH$_{4}$~\cite{Abe:2018-Zr} ($I4/mmm$)& 255 & 0.88 & 0.13 & 81$^a$ \\
MgH$_{4}$~\cite{Lonie:2012} ($I4/mmm$) & 100 & 0.74 & 0.13-0.1 & 29-37$^b$ \\
YH$_{4}$~\cite{Li:2015a} ($I4/mmm$) & 120 & 1.01 & 0.13-0.1 & 84-95$^a$\\
TbH$_{4}$~\cite{Hai:2021} ($I4/mmm$) & 200 & 0.70 & 0.13-0.1 & 31.3-41.3$^b$ \\
LaH$_4$~\cite{Liu:2017-La-Y} ($I4/mmm$) & 300 & 0.43 & 0.13-0.1 & 5-10$^a$, 5-10$^b$\\ 
ScH$_{7}$~\cite{Zurek:2018b} ($Cmcm$) & 300 & 1.84 & 0.1 & 213$^a$, 169$^b$  \\
AcH$_{16}$~\cite{Semenok-2018} ($P\bar{6}m2$) & 150 & 2.16 & 0.15-0.1 & 221-241$^a$, 171.3-199.2$^b$\\
\hline
\end{tabular}

\footnotesize{$^a$ $T_c$ was calculated using the Allen-Dynes modified McMillan equation (Equation~\ref{eq:ADM})}\\
\footnotesize{$^b$ $T_c$ was calculated by numerical solution of the Eliashberg equations}\\
\label{tab:mixed}}
\end{table*}

 Lower $T_c$s are associated with LaH$_4$~\cite{Liu:2017-La-Y}, ZrH$_4$~\cite{Abe:2018-Zr}, and ThH$_4$~\cite{Kvashnin:2018}. Even though La is a trivalent metal like Sc and Y whose $I4/mmm$ MH$_4$ structures exhibit moderate to high $T_c$s, the $T_c$ of LaH$_4$ is predicted to be an order of magnitude smaller. Two explanations have been provided. First, La has a higher mass than its trivalent congeners, which is associated with reductions in phonon-mediated $T_c$~\cite{Liu:2017-La-Y}. Another facet is related to the increase in H$_a$-H$_a$ distance -- and decrease in interaction strength as shown by the calculated crystal orbital Hamilton populations integrated to the Fermi level (-iCOHP) values between the H$_a$ atoms -- owing to La's large size~\cite{Bi:2021}. The low estimated $T_c$ in ThH$_4$ arises from a similar lack of H$_a$-H$_a$ interaction~\cite{Kvashnin:2018}. On the other hand, the H$_a$-H$_a$ distance in ZrH$_4$ is more in line with that in trivalent metal tetrahydrides (Zr would be expected to be tetravalent), and its lower $T_c$ is likely due to the increased mass of Zr decreasing the value of $\omega_{\text{ln}}$~\cite{Abe:2018-Zr}. In contrast to what is found for MgH$_4$, CaH$_4$ and SrH$_4$ develop only a small DOS at $E_F$ from pressure induced band overlap, and are therefore unlikely to be good superconductors.

Due to its ubiquity, we have thus far focused on the $I4/mmm$ ThCr$_2$Si$_2$ structure, whose proclivity towards superconductivity varies greatly based on the metallic element. Metal atoms need to donate sufficient electron density to weaken H$_a$-H$_a$ interactions -- but not too much, and they need to be large enough to stretch the H$_a$-H$_a$ distance - but not too much. Fully breaking covalent interactions results in an essentially hydridic lattice, while weakening them with partial occupancy of H$_2$ $\sigma^*$ bands results in a metallic lattice with sizable DOS at $E_F$, as in YH$_4$. Conversely, phases with very low electron transfer into the H$_2$ units, as in CaH$_4$, undergo metallization through pressure-induced broadening of H$^-$ and H$_2$ $\sigma^*$ bands. This situation, fairly common in metal hydride phases mixing molecular H$_2$ units with atomic H, tends to result in a low projected H DOS at $E_F$ and poor propensity towards superconductivity.

In addition to $I4/mmm$ ScH$_4$, the Sc-H system contains  another phase mixing atomic and molcular H$_2$ -- $Cmcm$ ScH$_7$. In this structure, where layers of H$_2$ units alternate with layers of Sc and atomic H (Figure~\ref{fig:mixed_struct}b), Sc donates electron density into the molecular H$_2$ $\sigma^*$ bands, as well as to the atomic H, stretching the H$_2$ distance to roughly to 0.96~\AA{} at 300~GPa. This results in a robust DOS at $E_F$ and, at 300~GPa, an estimated $T_c$ of 169~K~\cite{Zurek:2018b}. In the Ac-H system, the $P\bar{6}m2$ AcH$_{16}$ structure (Figure~\ref{fig:mixed_struct}c) has Ac coordinated by 12 atomic hydrogens and six H$_2$ units~\cite{Semenok-2018}, resulting in a moderate H-based DOS at $E_F$. Several other Ac-H binary phases also contain a mixture of atomic and molecular H, but they are nonmetallic. At 150~GPa, the $T_c$ of $P\bar{6}m2$ AcH$_{16}$ is estimated to be 241~K with a large $\lambda$=2.16, although a strong pressure dependence is expected as a result of its layered structure.

\subsection{Clathrate based hydrogenic lattices} \label{sec:other}

\begin{figure*}
\begin{center}
\includegraphics[width=2\columnwidth]{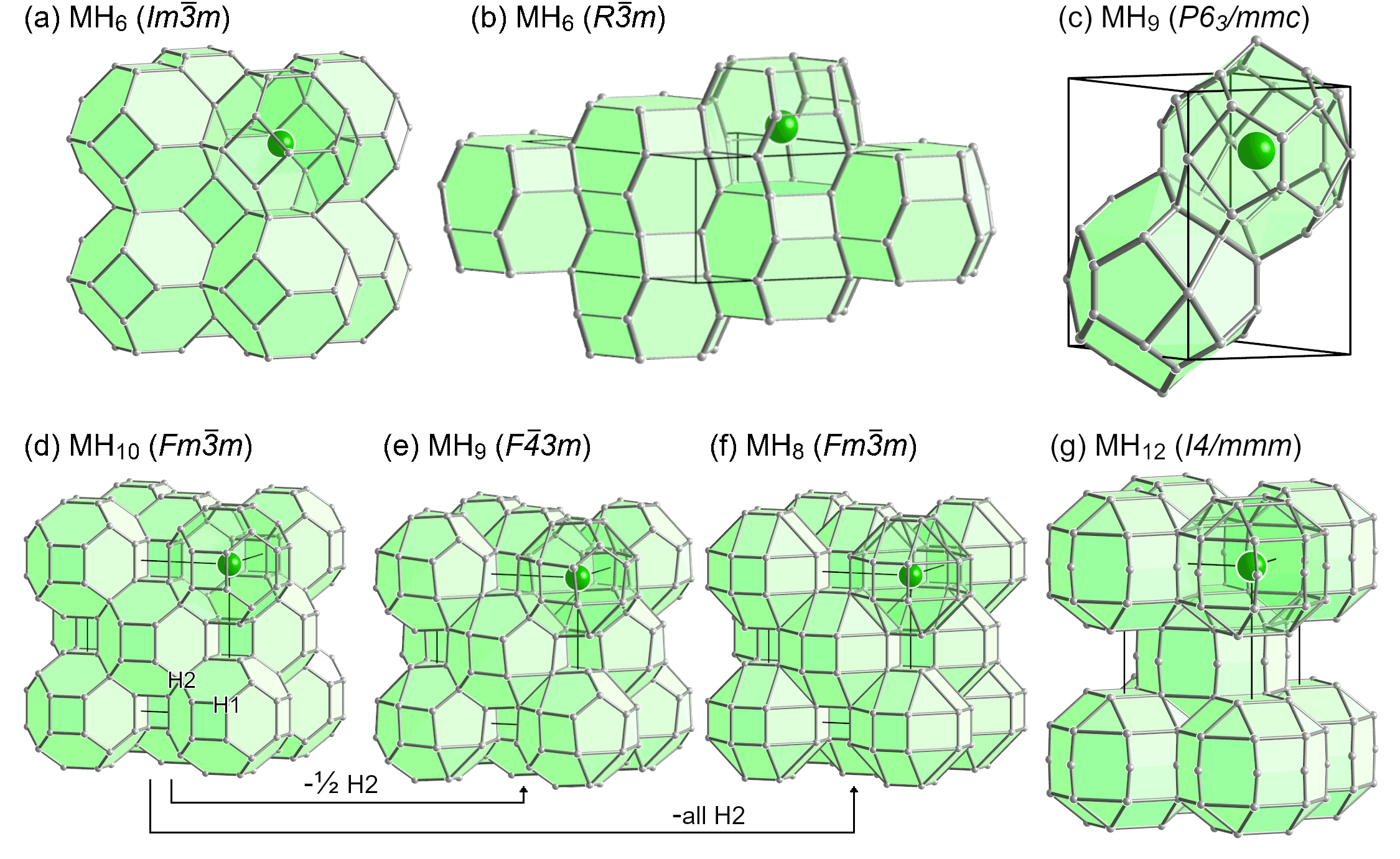}
\end{center}
\caption{Metal hydride structures with clathrate-like hydrogenic lattices. (a) $Im\bar{3}m$ symmetry structure adopted by several MH$_6$ phases with a body-centered lattice of M@H$_{24}$ polyhedra. (b) Distorted M@H$_{24}$ polyhedra comprising an $R\bar{3}m$ phase adopted by some MH$_6$ compounds with larger metal atoms. (c) $P6_3/mmc$ structure adopted by several MH$_9$ phases based on M@H$_{29}$ polyhedra. (d) $Fm\bar{3}m$ structure adopted by several MH$_{10}$ phases based on M@H$_{32}$ polyhedra with H1 (32\emph{f}) and H2 (8\emph{c}) sites. Removing half of the H2 atoms generates the (e) $F\bar{4}3m$ phase adopted by some MH$_9$ compounds, while removing all H2 atoms from the $Fm\bar{3}m$ MH$_{10}$ lattice yields (f) the $Fm\bar{3}m$ structure adopted by some MH$_8$ compounds. (g) $I4/mmm$ structure adopted by AcH$_{12}$ with similar coordination polyhedra to the $Fm\bar{3}m$ MH$_8$ structure where additional H atoms bisect eight of the square faces. }
\label{fig:clathrate_struct}
\end{figure*}

A calcium hydride compound consisting of Ca embedded in a curious ``sodalite-like'' hydrogenic cage~\cite{Wang:2012}, shown in Figure~\ref{fig:clathrate_struct}a, was the first in a long series of predicted phases featuring clathrate-like hydrogenic cages and remarkably high $T_{c}$s (Table~\ref{tab:clathrates}). In a great success for theory, the predictions for a number of members belonging to this class of structures have since been experimentally verified~\cite{Zulu:2018-La,Drozdov:2019,Snider:2021, Kong:2021a,Troyan:2021a,Chen:2021a}. In 2012, the aforementioned CaH$_{6}$ phase was predicted to be stable above 150~GPa in a crystal structure with $Im\bar{3}m$ symmetry~\cite{Wang:2012} soon found to be shared by several other high pressure MH$_{6}$ compounds, including MgH$_{6}$~\cite{Feng:2015a}, YH$_{6}$~\cite{Li:2015a,Peng:Sc-2017}, ScH$_{6}$~\cite{Peng:Sc-2017,Qian:Sc-2017,Zurek:2018b}, PuH$_{6}$~\cite{Zhao:2020}, TbH$_{6}$~\cite{Hai:2021}, EuH$_{6}$~\cite{Semenok:2021}, and (Nd-Lu)H$_{6}$~\cite{Sun:2020,Peng:Sc-2017}. 

This body-centered cubic structure features Ca atoms surrounded by a sodalite-like H$_{24}$ framework comprised of six square and eight hexagonal faces, denoted as [4$^{6}$6$^{8}$], with a square H$_{4}$ unit on each face of the conventional unit cell. At 150 GPa, the calculated H-H distances in CaH$_6$ measure 1.24 \AA, shorter than the H$_b$-H$_b$  distances (1.95 \AA) but longer than the H$_a$-H$_a$ distances (0.82 \AA) in $I4/mmm$ CaH$_{4}$ at the same pressure, suggestive of a weakly bound clathrate cage. The calculated electron localization function (ELF) provides further evidence for this weak bond between the H atoms, but no covalent bonds were present between Ca and H. In fact, the molecular orbital diagram predicted for a H$_{4}$ square unit possesses a half-occupied degenerate non-bonding state above the lowest-energy bonding state. If each Ca atom donated both its valence electrons to the hydrogen sublattice, each H could then possess up to 1/3 of an electron in excess, placed into this non-bonding state without having to populate an antibonding orbital. With the degenerate non-bonding state partially filled, Jahn-Teller distortions reducing the structural symmetry can lift this degeneracy, and phonon modes based on such symmetry reductions can contribute strongly to superconductivity by way of the EPC parameter. The calculated EPC of CaH$_6$ is remarkably large, $\lambda$=2.69 and $T_{c}$=220-235~K at 150~GPa, with most of the EPC being derived from symmetry breaking breathing and rocking motions within the H$_{4}$ units at $\Gamma$ as favored by the Jahn-Teller effect.

This discovery prompted numerous investigations into the possibility that other metal hydrides could possess similar structures when squeezed. 
In short order, a YH$_{6}$ phase isotypic to CaH$_{6}$ was predicted to be stable above 110 GPa when including ZPE effects \cite{Li:2015a}. 
With a calculated $\lambda$=2.93 at 120~GPa, the $T_{c}$ was estimated to be 251-264~K, even higher than the prediction for CaH$_{6}$. The shape of the Eliashberg spectral function was quite different from CaH$_{6}$; while the EPC in CaH$_{6}$ was found mostly to arise from T$_{2g}$ and E$_{g}$ modes at the $\Gamma$ point, the contributions to the EPC in YH$_{6}$ were much more dispersed throughout the phonon band structure.  
YH$_{6}$ has been synthesized at 160 GPa, where its $T_{c}$ was measured to be 224~K~\cite{Troyan:2021a}, lower than the initial theory predictions. The inclusion of anharmonic effects and setting $\mu^{*} $=0.19-0.22 were needed for the calculated $T_c$ to match up with experiment. 

It did not take long before theory predicted that MgH$_{6}$~\cite{Feng:2015a} and ScH$_{6}$~\cite{Peng:Sc-2017,Qian:Sc-2017,Zurek:2018b} $Im\bar{3}m$ phases would be stable above 263~GPa and 275~GPa, respectively. Similar to YH$_{6}$, the contributions to the EPC in MgH$_{6}$ were spread throughout the Brillouin zone. Several polymorphs of ScH$_{6}$ have been predicted, with symmetries including $Cmcm$~\cite{Zurek:2018b} and $P6_{3}/mmc$~\cite{Peng:Sc-2017,Qian:Sc-2017,Zurek:2018b}, suggesting complex phase behavior under pressure. Finally, recent explorations of the rare earth hydrides have identified Pu~\cite{Zhao:2020}, Tb~\cite{Hai:2021}, Eu~\cite{Semenok:2021}, and Pm-Lu~\cite{Sun:2020} analogues, with TbH$_{6}$ having an estimated $T_{c}$ up to 147~K at 300~GPa.

In addition to these, a number of other high-pressure MH$_{6}$ phases have been predicted with crystal structures that can be derived from distortions of the highly symmetric H$_{24}$ polyhedra that comprise the $Im\bar{3}m$ structure. One example is ZrH$_{6}$, which forms a phase with $I4/mmm$ symmetry above 275~GPa\cite{Abe:2018-Zr}, although it is very close in energy to a $P2{_1}/c$ structure until 320~GPa. This $I4/mmm$ structure is obtained by slightly stretching the cubic $Im\bar{3}m$ phase to a $c/a$ ratio of 1.08. More complex distortions occur in the $R\bar{3}m$ structure predicted for SrH$_{6}$, in which opposite hexagonal faces in the sodalite-like H$_{24}$ framework are opened up, resulting in a helical hydrogenic arrangement~\cite{Wang:2015a,Hooper:2013} (Figure~\ref{fig:clathrate_struct}b). Imaginary phonon modes calculated at low pressures involve the formation of H$_{2}\cdot\cdot\cdot$H$^{-}$ configurations along the helical chains and indicate possible liquid-like behavior as has been proposed in other hydride phases~\cite{Zaleski-Ejgierd:2011a,Hooper:2011a}. SrH$_{6}$ is a good metal, owing to equidistant 1-dimensional arrangements of H atoms with some electron donation from Sr, analogous to the electron transfer from Ca to H in CaH$_{6}$, resulting in a calculated $T_{c}$ of 156~K at 250~GPa~\cite{Zurek:2018d}. LaH$_{6}$ has also been predicted to form in this structure type at pressures of $\sim$100-150~GPa~\cite{Liu:2017-La-Y,Peng:Sc-2017}, although not necessarily to lie on the La-H convex hull. The H$_{24}$ cage is further broken up in an $Imm2$ structure predicted for BaH$_{6}$, in which H$_{3}^{-}$ units are present~\cite{Hooper:2012b}. Several possible BaH$_{6}$ geometries are nearly isoenthalpic over a range of pressures, with various hydridic, molecular, or H$_{3}^{-}$ motifs present in each.

Hydrogenic clathrate-like cages are also prevalent for the MH$_{9}$ stoichiometry. The predicted structure with the highest symmetry is a $P6_{3}/mmc$ arrangement of M@H$_{29}$ cages with a [4$^{6}$5$^{6}$6$^{6}$] framework (Figure~\ref{fig:clathrate_struct}c), and H-H distances slightly elongated relative to molecular H$_{2}$. Several phases have been predicted to assume this structure, including YH$_{9}$~\cite{Liu:2017-La-Y,Peng:Sc-2017}, ScH$_9$~\cite{Peng:Sc-2017}, CeH$_9$~\cite{Salke:2019, Sun:2020}, LaH$_9$~\cite{Peng:Sc-2017}, NdH$_9$~\cite{Sun:2020,Zhou:2020}, and (Pr-Er)H$_{9}$~\cite{Sun:2020}. $P6_{3}/mmc$ PrH$_{9}$~\cite{Zhou:2020} and EuH$_{9}$~\cite{Semenok:2021} have been identified in experiments, although in both cases a different structure was calculated to lie on the convex hull -- an $F\bar{4}3m$ geometry for PrH$_{9}$~\cite{Sun:2020,PenaAlvarez:2020} and a $P1$ Eu$_{4}$H$_{36}$ phases that arises from small distortions~\cite{Semenok:2021}.

\begin{table*}[!ht]
    \centering
    \def\arraystretch{1}
    \caption{Superconducting properties of select binary metal hydrides with clathrate-like hydrogen cages (Figure \ref{fig:clathrate_struct}).}
     \setlength{\tabcolsep}{3mm}{        
       \begin{tabular}{c c c c c c}
\hline
Structure & Pressure (GPa) & $\lambda$ & $\mu^*$ & $T_c$ (K) \\
(space group) & & & & \\
\hline
MgH$_6$~\cite{Feng:2015a} ($Im\bar{3}m$) & 300 & 3.29 & 0.12 & 263$^a$  \\
CaH$_6$~\cite{Wang:2012} ($Im\bar{3}m$) & 150 & 2.69 & 0.13-0.1 & 220-235$^b$  \\
ScH$_6$~\cite{Peng:Sc-2017} ($Im\bar{3}m$) & 300 & 1.2$^c$ & 0.13-0.1 & 90-100$^{b,c}$ \\
ScH$_6$~\cite{Abe:Sc-2017} ($Im\bar{3}m$) & 285 & 1.33 & 0.13-0.1 & 130-147$^a$ \\
ScH$_6$~\cite{Zurek:2018b} ($Im\bar{3}m$) & 350 & 1.25 & 0.1 & 135$^a$, 169$^b$ \\
YH$_6$~\cite{Li:2015a} ($Im\bar{3}m$) & 120 & 2.93 & 0.13-0.1 & 251-264$^b$ \\
YH$_6$~\cite{Peng:Sc-2017} ($Im\bar{3}m$) & 120 & 2.9$^c$ & 0.13-0.1 & 250-260$^{b,c}$ \\
YH$_6$~\cite{Heil:2019} ($Im\bar{3}m$) & 300 & 1.73 & 0.11 & 290$^d$ \\
SrH$_6$~\cite{Zurek:2018d} ($R\bar{3}m$) & 250 & 1.10 & 0.1 & 108$^a$, 156$^b$ \\
LaH$_6$~\cite{Peng:Sc-2017} ($R\bar{3}m$) & 100 & 1.9$^c$ & 0.13-0.1 & 155-165$^{b,c}$ \\
LaH$_6$~\cite{Semenok-2018} ($Im\bar{3}m$)$^e$ & 180 & 2.41 & 0.15-0.1 & 180-207$^a$, 217-235$^b$ \\
TbH$_6$~\cite{Hai:2021} ($Im\bar{3}m$) & 200 & 1.58 & 0.13-0.1 & 132.6-148.3$^a$ \\
ZrH$_{6}$~\cite{Abe:2018-Zr} ($I4/mmm$) & 295 & 1.20 & 0.13 & 114$^b$ \\
\hline
PaH$_8$~\cite{Xiao:2019} ($Fm\bar{3}m$)$^e$ & 10 & 1.58 & 0.13-0.1 & 71.4-79.4$^{a,e}$ \\
LuH$_8$~\cite{Sun:2020} ($Immm$) & 300 & 2.18 & 0.13-0.1 & 80.9-86.2$^a$ \\
LaH$_8$~\cite{Liu:2017-La-Y} ($C2/m$) & 300 & 1.12 & 0.13-0.1 & 114-131$^a$ \\
AcH$_{8}$~\cite{Semenok-2018} ($C2/m$) & 150 & 1.79 & 0.15-0.1 & 114.1-134$^a$, 134-149$^b$ \\
\hline
ScH$_9$~\cite{Peng:Sc-2017} ($P6_3/mmc$) & 400 & 1.3$^c$ & 0.13-0.1 & 170-180$^{b,c}$ \\
YH$_9$~\cite{Peng:Sc-2017} ($P6_3/mmc$) & 150 & 4.42 & 0.13-0.1 & 253-276$^b$ \\
LaH$_9$~\cite{Peng:Sc-2017} ($P6_3/mmc$) & 50 & 0.6$^c$ & 0.13-0.1 & 20-30$^{b,c}$ \\
CeH$_9$~\cite{Peng:Sc-2017} ($P6_3/mmc$) & 100 & 0.5$^c$ & 0.13-0.1 & 40-50$^{b,c}$ \\
CeH$_9$~\cite{Sun:2020} ($P6_3/mmc$) & 250 & 0.97 & 0.13-0.1 & 86.8-74.1$^a$ \\
CeH$_9$~\cite{Salke:2019} ($P6_3/mmc$) & 200 & 2.30 & 0.15-0.1 & 105-117$^a$ \\
YbH$_9$~\cite{Sun:2020} ($P6_3/mmc$) & 250 & 1.02 & 0.13-0.1 & 68.5-79.6$^a$ \\
CaH$_9$~\cite{Shao:2019} ($C2/m$) & 400 & 2.27 & 0.13-0.1 & 206.1-219.2$^a$, 260.6-285.2$^f$ \\
\hline
YH$_{10}$~\cite{Liu:2017-La-Y} ($Fm\bar{3}m$) & 250 & 2.56 & 0.13-0.1 & 305-326$^b$ \\
YH$_{10}$~\cite{Peng:Sc-2017} ($Fm\bar{3}m$) & 400 & 2.41 & 0.13-0.1 & 287-303$^b$ \\
YH$_{10}$~\cite{Heil:2019} ($Fm\bar{3}m$) & 300 & 2.41 & 0.11 & 310$^d$ \\
LaH$_{10}$~\cite{Liu:2017-La-Y} ($Fm\bar{3}m$) & 250 & 2.2 & 0.13-0.1 & 257-274$^b$ \\
LaH$_{10}$~\cite{Liu-2018-La} ($C2/m$) & 200 & 3.57 & 0.13-0.1 & 229-245$^a$ \\
LaH$_{10}$~\cite{Sun:2020} ($Fm\bar{3}m$) & 250 & 2.25 & 0.13-0.1 & 210.4-223.9$^a$ \\
CeH$_{10}$~\cite{Peng:Sc-2017} ($Fm\bar{3}m$) & 200 & 0.5$^c$ & 0.13-0.1 & 50-60$^{b,c}$ \\
CeH$_{10}$~\cite{Sun:2020} ($Fm\bar{3}m$) & 250 & 0.59 & 0.13-0.1 & 43.8-30.6$^a$ \\
YbH$_{10}$~\cite{Sun:2020} ($Fm\bar{3}m$) & 250 & 1.10 & 0.13-0.1 & 89.2-102.1$^a$ \\
TbH$_{10}$~\cite{Hai:2021} ($Fm\bar{3}m$) & 270 & 2.25 & 0.13-0.1 & 247.9-272.7$^{a,e}$ \\
TbH$_{10}$~\cite{Hai:2021} ($R\bar{3}m$) & 270 & 2.86 & 0.13-0.1 & 191.8-212.5$^a$ \\
AcH$_{10}$~\cite{Semenok-2018} ($R\bar{3}m$) & 200 & 3.46 & 0.15-0.1 & 177-204.1$^a$, 226-251$^b$ \\
CaH$_{10}$~\cite{Shao:2019} ($R\bar{3}m$) & 400 & 1.67 & 0.13-0.1 & 137.9-146.5$^a$, 156.9-175.3$^{e,f}$ \\
\hline
AcH$_{12}$~\cite{Semenok-2018} ($I4/mmm$) & 150 & 1.42 & 0.15-0.1 & 103.7-123.3$^a$, 148-173$^b$ \\
\hline
\end{tabular}

\footnotesize{$^a$ $T_c$ was calculated using the Allen-Dynes modified McMillan equation (Equation~\ref{eq:ADM})}\\
\footnotesize{$^b$ $T_c$ was calculated by numerical solution of the Eliashberg equations}\\
\footnotesize{$^c$ Values estimated from Figure 4 in Ref.~\citen{Peng:Sc-2017}}\\
\footnotesize{$^d$ Values calculated using fully anisotropic Migdal-Eliashberg theory as implemented in EPW~\cite{Margine:2013,Ponce:2016}}\\
\footnotesize{$^e$ Metastable at this pressure}\\
\footnotesize{$^f$ $T_c$ was calculated using the Allen-Dynes modified McMillan equation with strong coupling and shape corrections (Equations~\ref{eq:f1} and~\ref{eq:f2})}\\

\label{tab:clathrates}}
\end{table*}

$P6_3/mmc$ YH$_{9}$ (which lies on the convex hull above 200~GPa~\cite{Liu:2017-La-Y,Peng:Sc-2017}) has a remarkably large EPC constant, $\lambda$=4.42~at 150~GPa~\cite{Peng:Sc-2017}.  A synthesized yttrium superhydride with a measured $T_{c}$ of 262~K at ca.\ 180~GPa has been assigned as $P6_3/mmc$ YH$_9$~\cite{Snider:2021} on the basis of Raman spectroscopy, although a second experimental study found its $T_c$ to be roughly 20~K lower: 243~K at 201~GPa~\cite{Kong:2021a}. CeH$_{9}$ has an estimated $T_{c}$ of 74-87~K at 250~GPa~\cite{Sun:2020} and 105-117~K at 200~GPa~\cite{Salke:2019} but has also been experimentally observed at much lower pressures~\cite{Salke:2019,Li:2019-CeH9,Cui:2021}, where its $T_c$ was measured to be 57~K at 88~GPa increasing to 100~K at 130~GPa~\cite{Cui:2021}. Its shortest H-H distances at 100~GPa were 1.116~\AA, much shorter than in LaH$_{10}$, and in fact approaching the interatomic distances in atomic H~\cite{Salke:2019}. The markedly low pressure of 80~GPa at which CeH$_{9}$ was detected has been proposed to arise from particularly stong chemical precompression levied by the delocalized $4f$ electrons of Ce, suggesting that other rare earth hydrides with similar delocalized $f$ character may also be stabilized at relatively low pressures~\cite{Jeon:2020}. The low measured $T_c$ for (metastable) $P6_{3}/mmc$ PrH$_{9}$, < 9~K at 120~GPa~\cite{Zhou:2020}, was ascribed to the very small (<10\%) contribution from H states to the DOS at $E_F$, which was mostly comprised of Pr \emph{f} character. In fact, theoretical calculations suggest that many of the lanthanide MH$_9$ polyhydrides have a DOS at $E_F$ dominated by the M $f$ states, resulting in estimated $T_c$ values < 10~K~\cite{Sun:2020}.

As in the case of the MH$_{6}$ stoichiometry, MH$_{9}$ phases with structures based on distortions of the M@H$_{29}$ clusters have also been proposed. One example is a $C2/m$ symmetry CaH$_9$ structure stable above 235~GPa~\cite{Shao:2019}. H$_{2}$ and bent H$_{3}$ motifs comprise its hydrogenic network, reducing the overall symmetry of the H$_{29}$ cluster. Compared to $P6_{3}/mmc$ YH$_9$, in which three valence electrons are available for donation into the hydrogen network, Ca can only donate two, reflected in the calculated Bader charges. With fewer donated electrons occupying nonbonding or antibonding orbitals in the hydrogenic framework, the H$_2$ and H$_3$ units in $C2/m$ CaH$_9$ remain intact rather than disproportionating into the H$_{29}$ cluster present in YH$_9$ where the H-H contacts are nearly equidistant.  Finally, a TbH$_{9}$ phase with $C2/c$ symmetry was predicted to be stable above 175~GPa~\cite{Hai:2021}, although another study found only the $P6_{3}/mmc$ structure above 250~GPa~\cite{Sun:2020}.

Among the most promising candidates for high temperature superconductivity identified by CSP techniques are the LaH$_{10}$ and YH$_{10}$ phases, for which an $Fm\bar{3}m$ structure is predicted~\cite{Liu:2017-La-Y,Peng:Sc-2017,Sun:2020}. The hydrogenic network in this symmetry contains H$_{32}$ clusters with a [4$^{6}$6$^{12}$] framework (shown in Figure~\ref{fig:clathrate_struct}d). Another way to view these strctures are as metal atoms and H$_{8}$ hydrogen cubes comprising an interlocking fcc pattern. Hydrogen atoms occupy the 32\emph{f} (H1) and 8\emph{c} (H2) Wyckoff positions in the structure, with the H1 atoms forming square faces on the H$_{32}$ polyhedron, and the H2 atoms lying at the center of an (H1)$_{4}$ tetrahedron. The predicted $T_{c}$ values for LaH$_{10}$ and YH$_{10}$ were among the highest obtained for the metal clathrate superhydrides, 257-274~K for LaH$_{10}$ at 250~GPa~\cite{Liu:2017-La-Y}, and 305-326~K for YH$_{10}$ at 250 GPa~\cite{Liu:2017-La-Y}. Although the EPC constant is lower for YH$_{10}$ than YH$_{6}$ (2.56 vs. 2.93), the higher hydrogen content in YH$_{10}$ leads to a larger average phonon frequency, $\omega_\text{ln}$, and thus a very large expected $T_{c}$~\cite{Liu:2017-La-Y}. Because room temperature superconductivity is predicted in YH$_{10}$ its synthesis has been attempted, but so far only YH$_6$ and YH$_9$ have been made \cite{Snider:2021, Kong:2021a, Troyan:2021a}.  LaH$_{10}$ has successfully been synthesized~\cite{Geballe:2018a}, with a measured $T_{c}$ of 250~K at 170~GPa~\cite{Drozdov:2019} and 260~K  at 185~GPa~\cite{Zulu:2018-La}. 

Several additional REH$_{10}$ phases assuming $Fm\bar{3}m$ symmetry have been predicted, including CeH$_{10}$~\cite{Peng:Sc-2017,Li:2019,Sun:2020}, which has a maximum estimated $T_{c}$ of 156-168~K at 94~GPa~\cite{Li:2019} (although some studies also propose an $R\bar{3}m$ phase with a much lower estimated $T_{c}$ of <~56~K~\cite{Peng:Sc-2017}).  $Fm\bar{3}m$ CeH$_{10}$ has been observed experimentally with a measured $T_c$ of 115~K at 95~GPa~\cite{Cui:2021}. Alongside CeH$_9$, this reflects the impressive ability of cerium hydrides to maintain stability to -- for high-$T_c$ metal hydrides -- remarkably low pressures.   PrH$_{10}$~\cite{Peng:Sc-2017,Sun:2020}, PuH$_{10}$~\cite{Zhao:2020}, UH$_{10}$~\cite{Wang:2019}, and (Nd-Dy, Yb)H$_{10}$~\cite{Sun:2020} $Fm\bar{3}m$ phases have all been proposed to lie on the convex hull at various elevated pressures, and UH$_{10}$ has an estimated $T_{c}$ of 7-15~K at 550~GPa~\cite{Wang:2019}. An $Fm\bar{3}m$ phase is proposed to be near-isoenthalpic with an $R\bar{3}m$ structure for TbH$_{10}$; with estimated $T_{c}$s roughly 270~K above 230~GPa and 270~GPa for the $Fm\bar{3}m$ and $R\bar{3}m$ structures, respectively~\cite{Hai:2021}. 

Analogous to what is found for certain MH$_{6}$ and MH$_{9}$ phases, the $R\bar{3}m$ MH$_{10}$ structure can be derived by a distortion of the parent clathrate $Fm\bar{3}m$ phase where the hydrogenic network is broken up into puckered graphene-like nets. For the Ac-H~\cite{Semenok-2018} and Sr-H~\cite{Wang:2015a} systems the $R\bar{3}m$ configuration is predicted to be preferred, for which AcH$_{10}$ has an estimated  $T_{c}$ of 226-251~K at 200~GPa~\cite{Semenok-2018}. In contrast to CeH$_{10}$ and TbH$_{10}$, which have been predicted to assume both structure types, so far calculations have only located $R\bar{3}m$ CaH$_{10}$, which lies slightly above the convex hull at 400~GPa, with an estimated $T_{c}$ of 157-175~K~\cite{Shao:2019}.

These complex structural relationships encourage careful treatment of the energetics of competing phases, and the consideration of effects that go beyond the static lattice approximation. In the first reported synthetic study of LaH$_{10}$, an fcc lattice corresponding to the $Fm\bar{3}m$ structure was detected but then, with decompression, was observed to undergo a rhombohedral distortion~\cite{Geballe:2018a} to $C2/m$ symmetry (itself a distortion of the aforementioned $R\bar{3}m$ structure, with the La atoms maintaining $R\bar{3}m$ symmetry themselves). The estimated $T_{c}$ of this distorted phase was 229-245~K~\cite{Liu-2018-La}, a formidable albeit lower value than the predicted $T_{c}$ for the cubic phase (257-274~K~\cite{Liu:2017-La-Y}). Further theoretical work showed that the inclusion of quantum anharmonic effects had the dual effect of stabilizing the $Fm\bar{3}m$ phase over the competing less symmetric variants as well as extending the predicted pressure stability range for the $Fm\bar{3}m$ symmetry~\cite{Errea:2020}. It may be that these effects are also important in smoothing out the PES of the previously discussed MH$_6$, MH$_9$ and MH$_{10}$ stoichiometry phases that are distortions of the high-symmetry clathrates. Moreover, such quantum and anharmonic effects have been shown to suppress superconductivity in AlH$_{3}$~\cite{Hou:2021}, while quantum nuclear effects stabilize $Im\bar{3}m$ symmetry in the high-$T_c$ H$_3$S compound~\cite{Errea:2016}, highlighting their potential importance for the properties of metal hydrides.

Another family of structural derivatives of $Fm\bar{3}m$ MH$_{10}$ is generated by removing some or all of the 8\emph{c} H2 atoms leading to either an $F\bar{4}3m$ symmetry MH$_{9}$ (if half of the H2 atoms are removed, Figure~\ref{fig:clathrate_struct}e), or an $Fm\bar{3}m$ symmetry MH$_{8}$ (if all of the H2 atoms are removed, Figure~\ref{fig:clathrate_struct}f) stoichiometry. Two examples of the former are the cubic CeH$_{9}$~\cite{Salke:2019} and PrH$_{9}$~\cite{Sun:2020,PenaAlvarez:2020} phases expected to form instead of the $P6_{3}/mmc$ structure that was experimentally observed~\cite{Salke:2019,Zhou:2020,PenaAlvarez:2020}. $F\bar{4}3m$ PaH$_{9}$~\cite{Xiao:2019} and  (Pr-Sm,Ho,Yb)H$_{9}$~\cite{Sun:2020} phases have also been predicted. Examples of the latter include predicted $Fm\bar{3}m$ phases for (Nd, Ho-Tm)H$_{8}$~\cite{Sun:2020}, PaH$_{8}$~\cite{Xiao:2019}, PuH$_{8}$~\cite{Zhao:2020}, and UH$_{8}$~\cite{Wang:2019,Guigue:2020}, while LuH$_{8}$~\cite{Sun:2020} is expected to form in an $Immm$ tetragonally distorted version of the structure. At 300~GPa, LuH$_{8}$ has an estimated $T_{c}$ of 81-86~K~\cite{Sun:2020}, while PaH$_{8}$ reaches the convex hull at the remarkably low pressure of 32~GPa and maintains dynamic stability below 10~GPa, where its estimated $T_{c}$ is 71-79~K~\cite{Xiao:2019}. An additional $F\bar{4}3m$ PaH$_{5}$ stoichiometry is an even more extreme modification: every other H2@(H1)$_{4}$ tetrahedron of the $Fm\bar{3}m$ MH$_{10}$ structure is fully removed~\cite{Xiao:2019}. 

LaH$_{8}$~\cite{Liu:2017-La-Y}, AcH$_8$~\cite{Semenok-2018}, and YH$_{8}$~\cite{Liu:2017-La-Y} phases have also been predicted, although with $C2/m$ symmetry for La and Ac and $Cc$ symmetry for Y. Fragments of the $Fm\bar{3}m$ MH$_{10}$ and $Im\bar{3}m$ MH$_{6}$ geometries appear in LaH$_{8}$ and AcH$_8$. For even higher H content, an AcH$_{12}$ phase with $I4/mmm$ symmetry is metastable (by 0.05-0.06~eV/atom) in the 150-300~GPa pressure range~\cite{Semenok-2018}. Each Ac atom lies within a distorted rhombicuboctahedron of H atoms wherein one of the square faces has two additional H atoms along opposite edges (Figure~\ref{fig:clathrate_struct}g). A metastable RbH$_{12}$ phase with $Immm$ symmetry was found close to the convex hull at the impressively low pressure of 50~GPa, where it has an estimated $T_c$ of 115~K, increasing slightly to 126~K at 150~GPa~\cite{Hutcheon:2020}. High-frequency phonon modes involving motions of the hydrogen cage contributed strongly to the overall EPC of this structure. Notably, this study employed machine learning models to suggest high $T_c$ superconductors that were stable at lower pressures, illustrating the utility of machine learning methods in choosing promising systems.

\subsection{Other geometrical motifs} \label{sec:weird}

\begin{figure}
\begin{center}
\includegraphics[width=1\columnwidth]{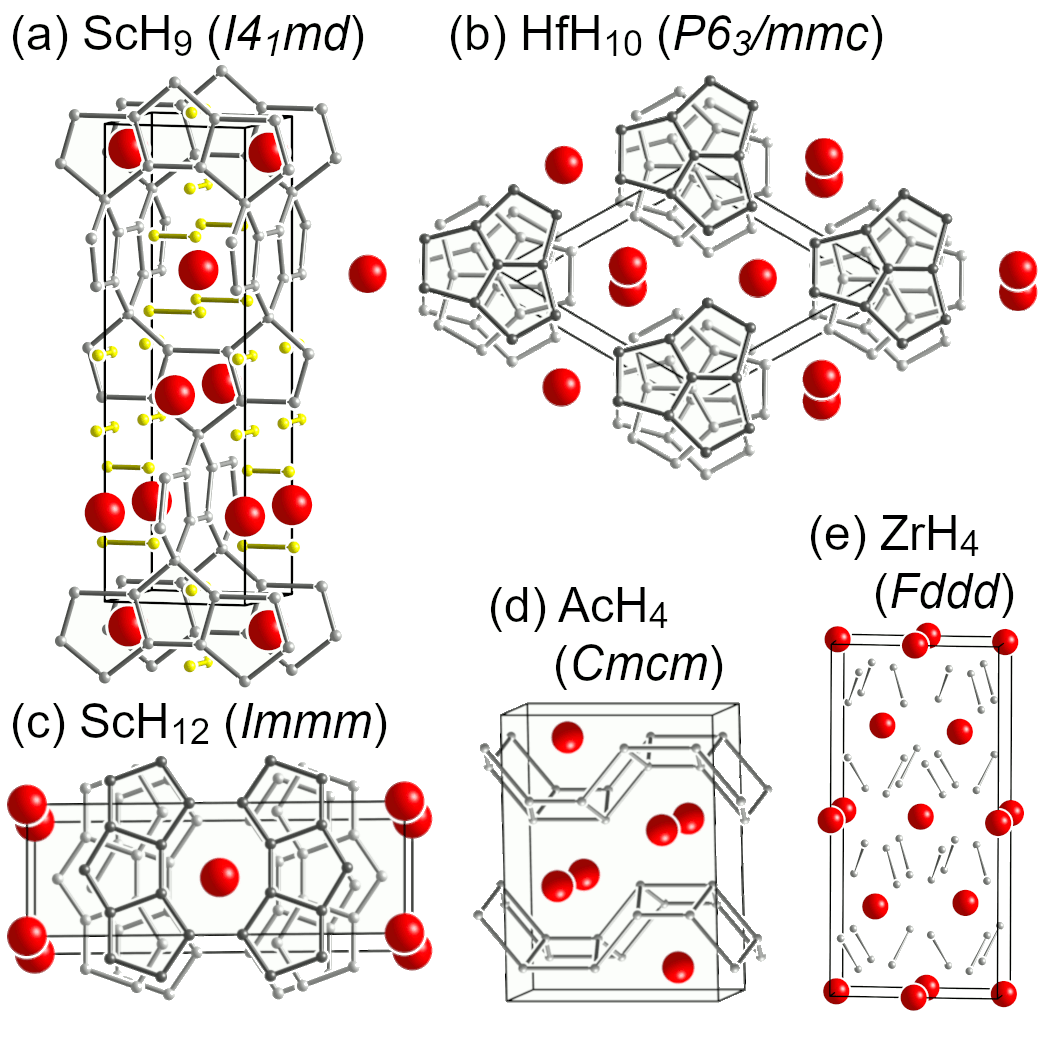}
\end{center}
\caption{Binary metal hydride structures featuring curious geometrical motifs. (a) Vertex-sharing perpendicularly oriented 1D chains of H$_5$ pentagons alongside molecular H$_2$ units in $I4_1md$ ScH$_9$. (b) Stacked triads of fused H$_5$ pentagons in $P6_3/mmc$ HfH$_{10}$. (c) 1D chains of H$_5$ pentagons in $Immm$ ScH$_{12}$. (d) Buckled nets of distorted squares in $Cmcm$ AcH$_4$. (e) Atomistic hydrogen in $Fddd$ ZrH$_4$ (shortest H-H contacts at a distance of 1.38 \AA~ are drawn to guide the eye).}
\label{fig:weird_struct}
\end{figure}

Most high-$T_c$ superconducting hydrides can be classified as possessing hydrogen lattices involving mixed atomic and molecular hydrogen (MgH$_4$), clathrate-like structures (LaH$_{10}$), or covalent 3D networks (H$_3$S), in which the electronic structure of the system is responsive to H atom motions, thereby conferring a strong EPC. Some phases, however, defy these classifications and feature curious hydrogenic motifs as well as superior superconducting properties (see Table \ref{tab:unique}). The family of scandium hydrides has been found to contain several of these strange hydrogen motifs alongside the clathrate-like ScH$_6$ and the ThCr$_2$Si$_2$-like ScH$_4$. Around 300~GPa, a ScH$_9$ phase with $I4_1md$ symmetry~\cite{Zurek:2018b} was found to be slightly lower in enthalpy than the previously predicted $P6_3/mmc$ symmetry clathrate ~\cite{Peng:Sc-2017}. In $I4_1md$ ScH$_9$, stacks of H$_5$ pentagons interspersed with H$_2$ molecular units are arranged along the $a$ and $b$ axes (Figure~\ref{fig:weird_struct}a). ELF plots clearly show bonding within the H$_5$ and H$_2$ units, and a fairly large EPC constant, $\lambda=1.94$, gives rise to an estimated $T_c$ of up to 233~K at 300~GPa.  

\begin{table*}[!ht]
    \centering
    \def\arraystretch{1}
    \caption{Superconducting properties of select binary metal hydrides with unique hydrogenic motifs (Figure \ref{fig:weird_struct}).}
     \setlength{\tabcolsep}{3mm}{        
       \begin{tabular}{c c c c c c}
\hline
Structure & Pressure (GPa) & $\lambda$ & $\mu^*$ & $T_c$ (K) \\
(space group) & & & & \\
\hline
ScH$_9$~\cite{Zurek:2018b} ($I4_1md$) & 300 & 1.23 & 0.1 & 163$^a$, 233$^b$ \\
ScH$_{10}$~\cite{Zurek:2018b} ($Cmcm$) & 250 & 1.17 & 0.1 & 120$^a$, 143$^b$ \\
ScH$_{10}$~\cite{Xie:2020} ($P6_3/mmc$) & 250 & 1.16 & 0.13-0.1 & 99-114$^{a,d}$, 112-124$^{b,e}$, 134-158$^f$  \\
HfH$_{10}$~\cite{Xie:2020} ($P6_3/mmc$)$^c$ & 250 & 2.77 & 0.13-0.1 & 152-167$^{a,d}$, 226-239$^{b,e}$, 213-234$^f$ \\
LuH$_{10}$~\cite{Xie:2020} ($P6_3/mmc$) & 200 & 1.36 & 0.13-0.1 & 105-118$^{a,d}$, 123-134$^{b,e}$, 134-152$^f$ \\
ZrH$_{10}$~\cite{Xie:2020} ($P6_3/mmc$)$^c$ & 250 & 1.77 & 0.13-0.1 & 151-167$^{a,d}$, 185-198$^{b,e}$, 199-220$^f$ \\
ScH$_{12}$~\cite{Zurek:2018b} ($Immm$) & 350 & 1.23 & 0.1 & 141$^a$, 194$^b$ \\
AcH$_{4}$~\cite{Semenok-2018} ($Cmcm$) & 100 & 0.89 & 0.15-0.1 & 43.2-60$^a$, 50-67$^b$ \\
ZrH$_{4}$~\cite{Abe:2018-Zr} ($Fddd$) & 140 & 1.08 & 0.13 & 78$^b$ \\
ZrH$_{6}$~\cite{Abe:2018-Zr} ($Cmc2_1$) & 215 & 0.92 & 0.13 & 70$^b$ \\
ZrH$_{6}$~\cite{Abe:2018-Zr} ($P2_1/c$) & 295 & 1.68 & 0.13 & 153$^b$ \\
\hline
\end{tabular}

\footnotesize{$^a$ $T_c$ was calculated using the Allen-Dynes modified McMillan equation (Equation \ref{eq:ADM})}\\
\footnotesize{$^b$ $T_c$ was calculated by numerical solution of the Eliashberg equations}\\
\footnotesize{$^c$ Metastable at this pressure}\\
\footnotesize{$^d$ $T_c$ was calculated using the Allen-Dynes modified McMillan equation with strong coupling and shape corrections (Equations~\ref{eq:f1} and~\ref{eq:f2})}\\
\footnotesize{$^e$ Matsubara-type linearized Eliashberg equations}\\
\footnotesize{$^f$ Gor'kov-Kresin formalism with $\lambda_\text{opt}$ and $\lambda_\text{ac}$ coupling constants to account for optical and acoustic phonon interactions with electrons}\\
\label{tab:unique}}
\end{table*}

Above 220~GPa, $Immm$ ScH$_{10}$ is predicted to become stable~\cite{Zurek:2018b}. Similar to ScH$_9$, this structure features H$_5$ pentagons, but here they are fused together in sets of three, with one pentagon fully closed and the other two slightly opened. Layers of the fused pentagons -- where bonding interactions are again revealed via plots of the ELF --  alternate with layers of Sc atoms. Essentially isoenthalpic with this structure is another ScH$_{10}$ geometry~\cite{Peng:Sc-2017,Xie:2020}  -- also based on triads of fused H$_5$ pentagons, but this time with $P6_3/mmc$ symmetry (Figure~\ref{fig:weird_struct}b). In this structure, the fused ``pentagraphene-like'' units are stacked along the $c$ axis and rotated relative to one another by 60\degree. It is predicted to lie only $\sim1-2$~meV/atom above the convex hull as HfH$_{10}$, $\sim18$~meV/atom above the convex hull as ZrH$_{10}$, and on the 300~GPa convex hull as LuH$_{10}$~\cite{Xie:2020}. Magnesium and thorium analogues -- chosen for their comparable electronegativities and atomic radii to Hf and Zr -- were found to be dynamically unstable, suggested to be due to insufficient electron transfer from Mg (which lacks $d$ electrons) and the too-large size of thorium.  The high $T_c$ values estimated for (Sc, Hf, Zr, Lu)H$_{10}$ (up to 213-234~K at 250~GPa for HfH$_{10}$) are credited to a large hydrogen-based DOS at $E_F$ and strong coupling between electrons and optical phonons. A Bader charge analysis indicates 0.13$e$ per H atom is transferred from Hf into antibonding H-based orbitals, lengthening H-H distances and increasing the H atom DOS at $E_F$. 

Pentagonal motifs surface again in $Cmmm$ ScH$_{12}$, with strips of edge-sharing H$_5$ pentagons running along the $c$ direction that are spaced by Sc atoms~\cite{Zurek:2018b} (Figure~\ref{fig:weird_struct}c). Short H-H distances between neighboring strips produce H$_8$ distorted octagons between them, with each Sc bracketed by one such H$_8$ octagon above and below. Such layered arrangements are reminiscent of the proposed high-pressure structure adopted by elemental H consisting of graphene-like sheets spaced by H$_2$ units, consistent with the especially high hydrogen content of this phase.

Between the remarkably low pressures of 75-100~GPa, $Cmcm$ AcH$_4$ is predicted to be stable~\cite{Semenok-2018}. It is based on corrugated layers composed of H in distorted square nets with Ac atoms in between (Figure~\ref{fig:weird_struct}d). At 100~GPa, its estimated $T_c$ extends up to a moderate 67~K~\cite{Semenok-2018}. An $Fddd$ structure for ZrH$_4$ consists of atomistic hydrogen (Figure~\ref{fig:weird_struct}e), transitioning to the previously discussed $I4/mmm$ structure above 210~GPa, while ZrH$_6$ with $Cmc2_1$ symmetry  -- involving a mixture of atomistic hydrogen and H@H$_3$ units in a slightly pyramidal configuration -- appears above 150~GPa until transitioning to either $I4/mmm$ or $P2_1/c$ symmetry at 275~GPa~\cite{Abe:2018-Zr}. The $P2_1/c$ structure, close in enthalpy to the tetragonally distorted $I4/mmm$ ZrH$_6$ clathrate structure and mostly containing atomistic H, has an estimated $T_c$ of 153~K at 295~K. 

\section{Ternary hydride materials} \label{sec:ternaries}

As described above, the phase diagrams of the various binary hydride systems have been thoroughly explored by theory, and experimental groups are making steady progress in synthesizing promising candidates. However, the high pressures required to stabilize these phases leave the quest for a room-temperature atmospheric pressure superconductor incomplete. One way to accomplish this goal is to once again make use of the concept of chemical precompression by adding a third element into the mix. This third element can slot into void spaces present in a binary hydride framework, function as a dopant to raise or lower electron counts, or facilitate the stabilization of altogether new geometrical motifs unobserved in binary hydrides. The vast number of possible two-element-plus-hydrogen combinations is at once promising and daunting. Computationally, building a ternary convex hull is highly intensive, with many more possible stoichiometries to explore. As such, the current literature on ternary superconducting hydride materials is at times piecemeal, with systematic trends not yet identified, limiting the researcher's predictive power with regard to these phases. In what follows, we present a literature survey of ternary superconducting hydride phases that have been predicted thus far, grouping them into families where possible based on structural features or relationship to known binary prototypes. For brevity, we focus on phases with predicted $T_c$s over ~100~K (see Table~\ref{tab:ternaries}).

\begin{figure}
\begin{center}
\includegraphics[width=1\columnwidth]{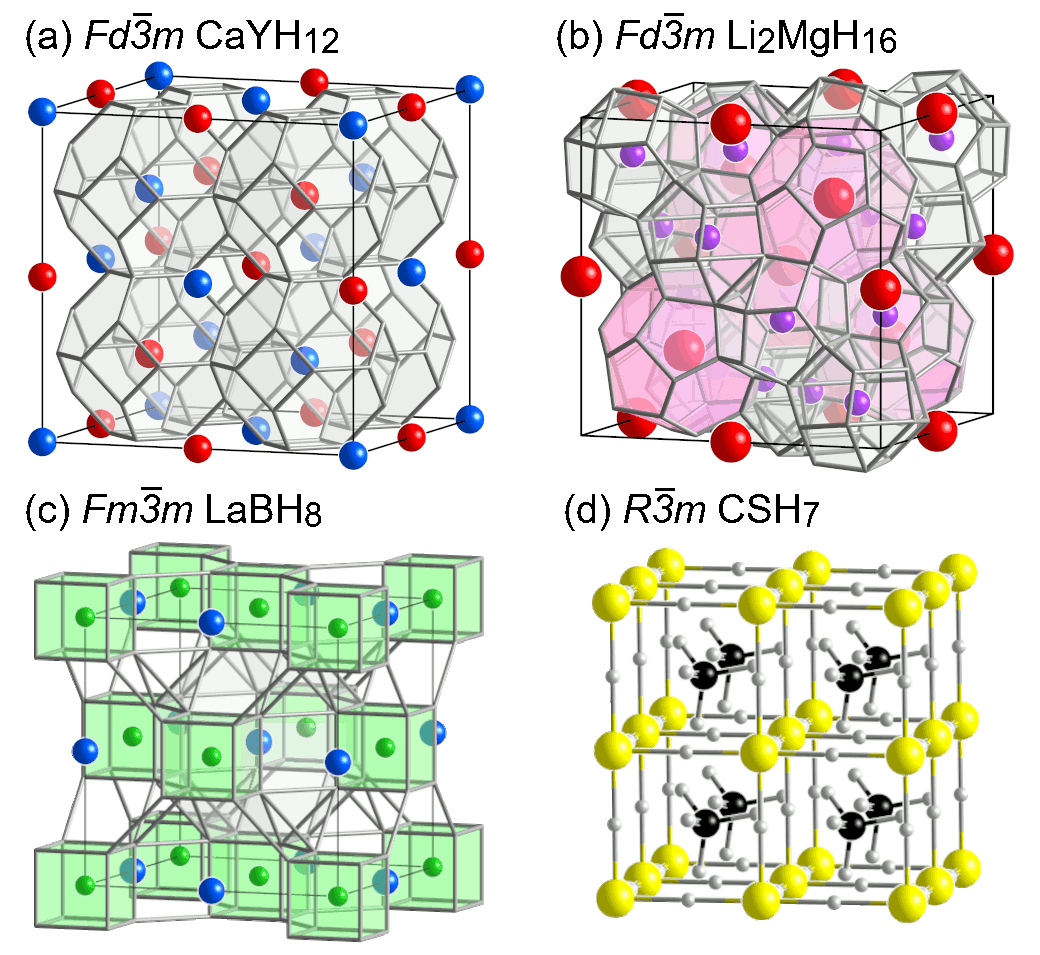}
\end{center}
\caption{Structures of selected ternary hydride superconducting phases. (a) CaYH$_{12}$ phase with $Fd\bar{3}m$ symmetry can be viewed as  a colored variant of the $Im\bar{3}m$ MH$_6$  structure. (b) $Fd\bar{3}m$ Li$_2$MgH$_{16}$ structure with interpenetrating Li@H$_{18}$ and Mg@H$_{28}$ clathrate lattices. (c) $Fm\bar{3}m$ LaBH$_8$ structure based on B intercalation into void spaces of an $Fm\bar{3}m$ MH$_8$ lattice. (d) $R\bar{3}m$ CSH$_7$ phase derived from methane substitution into cubic H$_3$S.}
\label{fig:Ternary_struct}
\end{figure}

Several ternary phases based on partial substitution of a third element in a binary hydride structure have been studied theoretically. One example is the CaYH$_{12}$ phase, which contains the same metal/hydrogen arrangement as found in $Im\bar{3}m$ CaH$_6$ and YH$_6$, but with the Ca and Y atoms in a B2 (CsCl-type) arrangement, reducing the overall symmetry to $Pm\bar{3}m$~\cite{Liang:2019}. Of course, with an even larger unit cell, more complex coloring patterns can occur. For CaYH$_{12}$, a structure with $Fd\bar{3}m$ symmetry, in which the Ca-Y sublattice displays a B1 (rock-salt) arrangement, possessed an even lower enthalpy and was dynamically stable above 170~GPa (Figure~\ref{fig:Ternary_struct}a). Although a full ternary convex hull was not constructed, the stability of these CaYH$_{12}$ phases was calculated relative to decomposition into CaH$_6$+YH$_6$ as well as CaH$_4$+YH$_3$+$\frac{5}{2}$H$_2$. The estimated $T_c$s of the two CaYH$_{12}$ phases were similar to those of the parent CaH$_6$ and YH$_6$ binaries. 

A similar series of Sc-Ca-H ternary phases based on Sc/Ca mixing in binary $I4/mmm$ MH$_4$ and $Im\bar{3}m$ MH$_6$ phases have also been proposed, with the phase having the highest estimated $T_c$ being $P4/mmm$ ScCaH$_8$ (212~K at 200~GPa)~\cite{Shi:2021}. Isotypic YBaH$_8$ is proposed to be stable to decomposition into BaH$_2$ and YH$_6$ above 50~GPa, with an estimated $T_c$ of 49.6~K at 50~GPa, which increases to 89.4~K at 150~GPa~\cite{Liu:2021b}. Machine learning investigations on high-$T_c$ materials created by substituting light elements such as Be, Li, Mg, and Na into YH$_x$ phases prompted further searches into the Y-Mg-H system~\cite{Song:2021b}. This led to a proposed metastable $Fd\bar{3}m$ YMgH$_{12}$ phase analogous to  $Fd\bar{3}m$ CaYH$_{12}$ having an estimated $T_c$ up to 189~K at 300~GPa~\cite{Song:2021}, as well as a second YMgH$_{12}$  phase with $Cmmm$ symmetry (based on Ca and Y atoms in alternating layers along the \emph{c} axis) with estimated $T_c$ up to 153~K at 250~GPa. 

With the extremely high $T_c$s of binary La and Y hydrides, a clear choice is to consider phases based on their mixture. A joint experimental/theoretical study~\cite{Semenok:2021b} demonstrated both synthetic feasibility and high $T_c$, measuring $T_c$s of 253~K and 237~K for (La/Y)H$_{10}$ and (La/Y)H$_{6}$ respectively, with the latter being higher than measured for binary YH$_6$. This result indicates the beneficial nature of La incorporation into the Y-H binary. The ternary convex hull was also calculated at 200~GPa. Notably, the only La-Y-H ternary that lay on it at 0~K was an $R\bar{3}m$ symmetry LaYH$_{20}$ structure, while increases in temperatue eventually found La$_2$YH$_{12}$, La$_4$YH$_{30}$, LaYH$_{12}$, La$_4$YH$_{50}$, LaY$_{4}$H$_{50}$, and La$_2$YH$_{18}$ (specific La/Y fractions for (La/Y)H$_4$, (La/Y)H$_6$, and (La/Y)H$_{10}$) all present on the convex hull, highlighting the importance of thermal effects and reinforcing the importance of considering metastable phases from the 0~K convex hull.

The above phases can all be constructed by partially substituting a third element onto one (or more) lattice sites in a known binary phase. However, as noted above, ternaries can also be created by adding a third element into void regions in a binary phase, thereby further tuning doping and leveraging the idea of ``chemical precompression.'' Two particularly fascinating materials based on this strategy are Li$_2$MgH$_{16}$, an electron-doped variant of MgH$_{16}$, and LaBH$_8$, a cousin of the high-$T_c$ binary LaH$_{10}$. 

Binary MgH$_{16}$, based on a network of Mg$^{2+}$ cations in a lattice of H$_2^{\delta-}$ molecules, has been predicted to lie on the convex hull above 200~GPa~\cite{Lonie:2012}. The donated electron density from the Mg atoms occupies the H$_2$ $\sigma^*$ orbital, weakening the H$_2$ units -- although not to the extent of forming fully extended networks as in the clathrate structures. The addition of lithium as an electron-doping agent was proposed to further perturb the H$_2$ units, resulting in the $P\bar{3}m1$ Li$_2$MgH$_{16}$ structure, which lies only 20~meV/atom shy of the convex hull at 300~GPa.~\cite{Sun:2019} In this structure, the electrons from Li destabilize the H-H bonds even further. An $Fd\bar{3}m$ phase, which overtakes the $P\bar{3}m1$ structure above 453~GPa, has fully dissociated the H$_2$ units to form clathrate cages (Figure~\ref{fig:Ternary_struct}b). This structure has a very large hydrogen-dominant electronic DOS at $E_F$. Because the estimated EPC parameter, $\lambda$=3.35, yields a $T_c$ of 351~K at 300~GPa, this phase has been coined a ``hot'' superconductor~\cite{Sun:2019}. The structure remains dynamically stable down to 250~GPa, where its estimated $T_c$ increases further to 430-476~K. Here, electron doping through the addition of a third element changes the structural and electronic landscape of the phase, enhancing its superconducting properties.

Despite these substantial changes, adding Li to MgH$_{16}$ does not drastically alter the pressure range of phase stability. The situation is somewhat different for LaBH$_{8}$, where boron atoms occupy H$_8$ cubes that are empty in structurally related LaH$_{10}$ (Figure~\ref{fig:Ternary_struct}c).~\cite{diCataldo:2021} Although it does not lie on the La-B-H ternary convex hull until 110~GPa, this phase maintains dynamic stability down to an impressive 40~GPa, suggesting that synthesis above 100~GPa then decompression could be a viable strategy to quench this phase to lower pressures. In contrast, LaH$_{10}$ is not dynamically stable at pressures below 210~GPa.~\cite{Liu:2017-La-Y} Furthermore, the estimated $T_c$ of LaBH$_8$ increases with decreasing  pressure, going from ca.\ 42~K at 100~GPa up to ca.\ 126~K at 50~GPa.~\cite{diCataldo:2021} This remarkable stability at low pressures is linked to the addition of boron atoms providing additional ``chemical precompression'' to the hydrogen lattice, thereby increasing the effective pressure even at lower external pressures. An additional study found that LaBeH$_8$ (in the same structure type) was dynamically stable down to 50~GPa, where its $T_c$ was estimated to be 191~GPa~\cite{Zhang:2021}. In addition to the La-B-H and La-Be-H structures, LaAlH$_8$, CaBH$_8$, and YBeH$_8$ were also found to be be dynamically stable at pressures below 100~GPa and LaBeH$_8$, CaBH$_8$, YBeH$_8$, and SrBH$_8$ (dynamically stable above 210~GPa) were all found to lie on the convex hull~\cite{Zhang:2021}. Finally, whereas the previous XYH$_8$ phases involve placement of Be or B into H$_8$ cubes in an MH$_8$ hydrogen framework, another possibility for this structure is presented by KB$_2$H$_8$~\cite{Gao:2021} and LaC$_2$H$_8$~\cite{Durajski:2021}, in which B and C are respectively placed into the H$_4$ tetrahedral vacancies of an MH$_8$ structure, resulting in a ``colored'' version of the LaH$_{10}$ lattice. KB$_2$H$_8$ remains dynamically stable down to 12~GPa, where its $T_c$ is estimated to be 134-146~K~\cite{Gao:2021}. LaC$_2$H$_8$ is dynamically stable above 70~GPa, with an estimated $T_c$ of 69~K, increasing to 140~K at 250~GPa~\cite{Durajski:2021}. As such, the XYH$_8$ and XY$_2$H$_8$ families, with a hydrogenic lattice stabilized to low pressure by scaffolding atoms, could be a useful vector for further investigations. 

The B@H$_8$ cubes featured in several of the aforementioned phases are one instance of B@H$_n$ polyhedra prevalent in ternary superconducting systems. Other examples are B@H$_6$ octahedra found in Li$_2$BH$_6$~\cite{Kokail:2017}, CaBH$_6$, and Ca$_2$B$_2$H$_{13}$~\cite{diCataldo:2020}. In the former, the B@H$_6$ octahedra are arranged in an fcc lattice intercalated with Li atoms, resulting in an $Fm\bar{3}m$ structure that is stable down to 100~GPa. Its $T_c$ has been estimated to be 98~K at 100~GPa, with most contributions arising from atomic vibrations of intermediate frequencies. CaBH$_6$ with $Pa\bar{3}$ symmetry is metastable by ca.\ 100~meV/atom above 100~GPa, while Ca$_2$B$_2$H$_{13}$ has $Pm$ symmetry and lies on the convex hull above 275~GPa. At 300~GPa, the estimated $T_c$ of CaBH$_6$ is 97-119~K, although little pressure dependence is observed down to 100~GPa, while that of Ca$_2$B$_2$H$_{13}$ is 75-89~K at 300~GPa. Since they lie closer to $E_F$ than in other alkali and alkaline earth metals, the interactions -- or lack thereof -- of the Ca \emph{d} orbitals with the rest of the atoms are postulated to be key to the superconducting properties of the Ca-B-H phases. A metastable LaBH$_{17}$ stoichiometry phase was identified alongside LaBH$_8$~\cite{diCataldo:2021b}, with La atoms encapsulated in clathrate-like H$_{32}$ cages interspersed with B$_2$H$_{10}$ units.

\begin{table*}[!ht]
    \centering
    \def\arraystretch{1}
    \caption{Superconducting properties of select ternary hydrides (Figure \ref{fig:Ternary_struct}).}
     \setlength{\tabcolsep}{3mm}{        
       \begin{tabular}{c c c c c c}
\hline
Structure & Pressure (GPa) & $\lambda$ & $\mu^*$ & $T_c$ \\
(space group) & & & & \\
\hline
CaYH$_{12}$~\cite{Liang:2019} ($Pm\bar{3}m$)$^e$ & 200 & 2.06 & 0.13-0.1 & 196-215$^{a,f}$, 233-248$^b$ \\
CaYH$_{12}$~\cite{Liang:2019} ($Fd\bar{3}m$) & 200 & 2.20 & 0.13-0.1 & 207-226$^{a,f}$, 243-258$^b$ \\
ScCaH$_8$~\cite{Shi:2021} ($P4/mmm$) & 160 & 1.62 & 0.1 & 130.2$^{a,f}$ \\
ScCaH$_8$~\cite{Shi:2021} ($P4/mmm$) & 200 & 1.89 & 0.1 & 212.1$^{a,f}$ \\
ScCaH$_{12}$~\cite{Shi:2021} ($Pm\bar{3}m$) & 160 & 1.80 & 0.1 & 174.9$^{a,f}$ \\
ScCaH$_{12}$~\cite{Shi:2021} ($Pm\bar{3}m$) & 300 & 1.68 & 0.1 & 193.2$^{a,f}$ \\
YMgH$_{8}$~\cite{Song:2021} ($P4/mmm$) & 300 & 1.63 & 0.13-0.1 & 115-125$^a$ \\
YMgH$_{12}$~\cite{Song:2021} ($Fd\bar{3}m$)$^e$  & 300 & 1.81 & 0.13-0.1 & 175-189$^a$ \\
YMgH$_{12}$~\cite{Song:2021} ($Cmmm$) & 250 & 1.66 & 0.13-0.1 & 141-153$^a$ \\
La$_2$YH$_{18}$~\cite{Semenok:2021b} ($Pm\bar{3}m$) & 180 & 2.74$^c$ & 0.15-0.1 & 197-229$^a$, 248-270$^b$ \\
LaYH$_{12}$~\cite{Semenok:2021b} ($Pm\bar{3}m$) & 180 & 2.82 & 0.15-0.1 & 176-203$^a$, 223-241$^b$ \\
La$_4$YH$_{30}$~\cite{Semenok:2021b} ($Pm\bar{3}m$) & 180 & 2.68$^c$ & 0.15-0.1 & 206-237$^a$, 245-265$^b$ \\
LaYH$_{20}$~\cite{Semenok:2021b} ($R\bar{3}m$) & 180 & 3.87 & 0.15-0.1 (0.2) & 232-266$^a$, 281-300 (266)$^b$ \\
Li$_2$MgH$_{16}$~\cite{Sun:2019} ($Fd\bar{3}m$)$^e$ & 250 & 4.0 & 0.16-0.1 & 430-473$^b$ \\
LaBH$_8$~\cite{diCataldo:2021} ($Fm\bar{3}m$) & 50 & 1.54 & 0.09 & 122$^b$, 126$^{b,d}$\\
LaBH$_8$~\cite{Liang:2021} ($Fm\bar{3}m$) & 50 & 2.29 & 0.13-0.1 & 102-108$^a$, (154, $\mu^*=0.1$)$^b$\\
LaBH$_{17}$~\cite{diCataldo:2021b} ($I222$)$^e$ & 300 & 3.3 & 0.1 & 180$^b$ \\
KB$_2$H$_8$~\cite{Gao:2021} ($Fm\bar{3}m$) & 12 & 2.99 & 0.15-0.1 & 134-146$^b$ \\
LaC$_2$H$_8$~\cite{Durajski:2021} ($Fm\bar{3}m$) & 250 & 1.63 & 0.1 & 107$^a$, 140$^b$ \\
Li$_2$BH$_6$~\cite{Kokail:2017} ($Fm\bar{3}m$) & 100 & 0.94 & 0.1 & 98$^a$ \\
CaBH$_6$~\cite{diCataldo:2020} ($Pa\bar{3}$) & 100 & 1.93 & 0.15-0.1 & 100-114$^a$ \\
Ca$_2$B$_2$H$_{13}$~\cite{diCataldo:2020} ($Pm$) & 300 & 1.37 & 0.15-0.1 & 75-89$^a$ \\
H$_6$SSe~\cite{Liu:2018} ($Pm\bar{3}m$) & 200 & 1.76 & 0.1 & 196$^{b}$ \\
H$_6$SSe~\cite{Liu:2018} ($Cmmm$) & 200 & 1.60 & 0.1 & 181$^{b}$ \\
H$_6$SSe~\cite{Liu:2018} ($Fd\bar{3}m$) & 200 & 0.99 & 0.1 & 115$^{b}$ \\
H$_3$S$_{(1-x)}$P$_x$ $^{g,}$~\cite{Ge:2016} ($Im\bar{3}m$) & 250 & 2.44 & 0.1 & 280$^{a,f}$ \\
H$_3$S$_{(1-x)}$C$_x$ $^{g,}$~\cite{Ge:2020} ($Im\bar{3}m$) & 260 & 2.43$^h$ & 0.15-0.1 & 289$^{a,f}$ \\
H$_3$S$_{(1-x)}$Si$_x$ $^{g,}$~\cite{Ge:2020} ($Im\bar{3}m$) & 230 & 2.46$^h$ & 0.15-0.1 & 283$^{a,f}$ \\
CSH$_7$ ~\cite{Sun:2020b} ($I\bar{4}3m$) & 100 & 3.64 & 0.13-0.1 & 114-126$^a$, 173-181$^b$\\
CSH$_7$ ~\cite{Cui:2020} ($R\bar{3}m$) & 150 & 2.47 & 0.13-0.1 & 143-152$^a$, 181-194$^b$\\
CSH$_7$ ~\cite{Cui:2020} ($Pnma$) & 150 & 3.06 & 0.13-0.1 & 116-122$^a$, 157-170$^b$\\
CSH$_7$ ~\cite{Cui:2020} ($Cm$) & 100 & 1.20 & 0.13-0.1 & 86-98$^a$, 97-108$^b$\\
H$_6$SSe~\cite{Liu:2018} ($Pm\bar{3}m$) & 200 & 1.76 & 0.1 & 246$^b$ \\
H$_6$SSe~\cite{Liu:2018} ($Cmmm$) & 200 & 1.60 & 0.1 & 196$^b$ \\
H$_6$SSe~\cite{Liu:2018} ($Fd\bar{3}m$) & 200 & 0.99 & 0.1 & 115$^b$ \\
CaSH$_3$ ~\cite{Yan:2020} ($P\bar{6}m2$) & 128 & 1.64 & 0.13-0.1 & 71-78$^a$, 94-101$^b$ \\
MgCH$_4$~\cite{Tian:2015} ($P4/nmm$) & 105 & 1.49$^i$ & 0.1 & 121$^a$ \\
LiPH$_6$~\cite{Shao:2019} ($Pm\bar{3}$) & 200$^e$ & 1.62 & 0.13-0.1 & 125.3-136.5$^a$, 149.6-167.3$^{a,f}$ \\
LiP$_2$H$_{14}$ ~\cite{Li:2020} ($R\bar{3}$) & 230 & 1.69 & 0.16-0.1 & 116-143$^a$, 138-169$^b$ \\
NaP$_2$H$_{14}$~\cite{Li:2020} ($R\bar{3}$) & 400 & 1.45 & 0.1 & 111-139$^a$, 130-159$^b$ \\
NSiH$_{11}$~\cite{Liu:2021} ($P2_1/m$) & 300 & 1.29 & 0.13-0.1 & 98-110$^a$ \\

\hline
\end{tabular}

\footnotesize{$^a$ $T_c$ was calculated using the Allen-Dynes modified McMillan equation (Equation \ref{eq:ADM})}\\
\footnotesize{$^b$ $T_c$ was calculated by numerical solution of the Eliashberg equations}\\
\footnotesize{$^c$ Values obtained from the interpolation method}\\
\footnotesize{$^d$ Values calculated using fully anisotropic Migdal-Eliashberg theory}\\
\footnotesize{$^e$ Metastable at this pressure}\\
\footnotesize{$^f$ $T_c$ was calculated using the Allen-Dynes modified McMillan equation with strong coupling and shape corrections (Equations~\ref{eq:f1} and~\ref{eq:f2})}\\
\footnotesize{$^g$ Calculated within the virtual crystal approximation (VCA)}\\
\footnotesize{$^h$ Values estimated from Figure 2 in Ref.\citen{Ge:2020} }\\
\footnotesize{$^i$ Values estimated from Figure 4 in Ref.\citen{Tian:2015} }\\
\label{tab:ternaries}}
\end{table*}

A third fruitful family of ternary hydrides are those based on covalent networks such as those found in H$_3$S -- the first high $T_c$ hydride to be synthesized.~\cite{Drozdov:2015a} Key to its record breaking superconducting properties is the presence of a pair of van Hove singularities (vHs) bracketing $E_F$. Since doping H$_3$S could shift $E_F$ to place it more directly at the peak in the DOS created by the vHs -- increasing the EPC and concomitant $T_c$ -- several studies have investigated the effect of adding a third element into the H$_3$S framework. One of these examined the effects of replacing half of the sulfur atoms with selenium, finding three dynamically stable H$_6$SSe structures at 200~GPa, with $Pm\bar{3}m$, $Cmmm$, and $Fd\bar{3}m$ symmetries arising from different S/Se substitution patterns~\cite{Liu:2018}. At this pressure, their $T_c$s were predicted to be 196, 181, and 115~K respectively, all lower than the 246~K estimated for $Im\bar{3}m$ H$_3$S at 200~GPa. Other thermodynamically stable S/Se compositions were explored using the cluster expansion method, all of which had greatly reduced estimated $T_c$s compared to cubic H$_3$S and H$_3$Se~\cite{Amsler:2019}.

Considering other ternary H$_3$S-based systems, studies using the virtual crystal approximation (VCA)~\cite{Ge:2016,Ge:2020,Hu:2020}  found that small amounts of phosphorus, carbon or silicon doping could shift $E_F$ sufficiently to increase the $T_c$ of the model H$_3$S$_{(1-x)}$Z$_x$ phase to 280~K (Z=P)~\cite{Ge:2016}, 289~K (Z=C)~\cite{Ge:2020} and 283~K (Z=Si)~\cite{Ge:2020}. In fact, a ``carbonaceous sulfur hydride'' phase was recently reported to be the world's first room-temperature superconductor~\cite{Snider:2020}, although the conclusions of the manuscript have been called into question~\cite{Hirsh:2021,Dogan:2021,Wang:2021}. Thus far, the C-S-H phase has eluded characterization, given the experimental difficulty of obtaining structural information for light element systems and in particular those under pressure. Further complicating the matter, the aforementioned VCA results do not account for the local bonding and structural modifications that result from substituting an actual dopant atom into the H$_3$S lattice~\cite{Wang:2021}, and further exploration is required -- and ongoing -- to determine the precise structural motifs within the carbonaceous sulfur hydride superconductor~\cite{Guan:2021,Wang:2021b}.

Prior to the experimental observation of the carbonaceous sulfur hydride system, two groups identified promising metastable C-S-H phases based on methane intercalation into an H$_3$S lattice~\cite{Cui:2020,Sun:2020b} (Figure~\ref{fig:Ternary_struct}d). The CSH$_7$ stoichiometry was especially promising, and various dynamically stable phases that differed via rotation of the methane molecules were identified. The phases with the highest predicted $T_c$s were $I\bar{4}3m$ CSH$_7$ (181~K at 100~GPa~\cite{Sun:2020b}) and $R\bar{3}m$ CSH$_7$ (181-194~K at 150~GPa~\cite{Cui:2020}). However, it is uncertain if these structural motifs are present in the C-S-H phase synthesized in Ref.\ \citen{Snider:2020}. For example, recent experimental investigations~\cite{Bykova:2021,Lamichhane:2021} suggest the formation of an Al$_2$Cu-based structure within the C-S-H system under pressure, rather than a phase derived from the $Im\bar{3}m$ H$_3$S structure.

One of the most well-known structures among intermetallic superconductors is the A15 phase adopted by Nb$_3$Sn~\cite{Matthias:1954} ($T_c$=18~K). A ternary $Pm\bar{3}$ LiPH$_6$ phase comprised of a colored A15 lattice was predicted above 250~GPa~\cite{Shao:2019}. A large peak in the DOS just below $E_F$ was observed, with an estimated $T_c$ as high as 167~K at 200 GPa. ELF analysis indicated that the H atoms are not bound to one another, with H-H distances of 1.14~\AA{} at 300~GPa.  In addition to a large H-based contribution to the DOS at $E_F$, the low atomic mass of Li was also thought to enhance $T_c$, especially upon comparison with previously identified isotypic MgSiH$_6$ and MgGeH$_6$ phases with much lower $T_c$s (63~K~\cite{Ma:2017b} and 67~K~\cite{Ma:2017}). Additional investigation of the Li-P-H system showed that the inclusion of ZPE effects renders several phases including LiPH$_6$ metastable. From 250-400~GPa, a $R\bar{3}$ phase with LiP$_2$H$_{14}$ stoichiometry was found, built from P@H$_9$ polyhedra and interstitial Li and H atoms~\cite{Li:2020}. At 230~GPa this phase reaches an estimated $T_c$=169~K, with the majority of the EPC resulting from H-H stretching modes. Without performing full CSP searches, Na and Be were substituted for Li in this structure to examine the role of atomic mass on $T_c$, finding enhancement at high pressures with NaP$_2$H$_{14}$ having $T_c$=159~K at 400~GPa (relative to 90~K for LiP$_2$H$_{14}$ at this pressure).

With the high $T_c$s of the CaH$_6$ and H$_3$S phases in mind, the Ca-S-H system was investigated. In addition to a number of semiconducting phases based on H$_2$ molecules in a Ca/S lattice, a $P\bar{6}m2$ CaSH$_3$ structure with stacked alternating S-H honeycomb nets and CaH$_2$ layers was found to lie 3-28 meV/atom above the convex hull in the 150-250~GPa pressure range, and on the hull at 300~GPa~\cite{Yan:2020}. Two vHs were found to lie bracket $E_F$ resulting in a $T_c$ as high as 94-101~K at 128~GPa. It was proposed that replacing Ca by other alkaline earth metals, and S with heavier chalcogens could lead to other superconducting materials in this structure type.

In analogy to the chemical precompression strategy to metallize hydrogen, Mg-doping was proposed as a way to metallize methane, resulting in the prediction of a $P4/nmm$-symmetry MgCH$_4$ phase stable above 75~GPa, where it overtakes a semiconducting $Pban$ structure~\cite{Tian:2015}. A maximum $T_c$ of 121~K was predicted at 105~GPa.

With the aim of incorporating large amounts of H into the system, prompted by the report of NH$_4$ and NH$_7$ structures~\cite{Song:2019}, N-Si-H ternaries were explored at 300~GPa~\cite{Liu:2021}. While all candidate ternary structures were metastable, the lowest-lying of these was $P2_1/m$ NSiH$_{11}$, lying 42~meV/atom above the convex hull. Weakly bound H$_2$ units account for most of the hydrogen in the phase, with some atomic hydrogen covalently linked to nitrogen. At 300~GPa, NSiH$_{11}$ has an estimated $T_c$ of 98-110~K, with the interaction of N atoms with the Si-H network leading to almost half of the EPC.

\section{Conclusion and Outlook} \label{sec:outlook}
Less than a decade after the theoretical prediction of high temperature superconductivity with a $T_c$ up to 220-235~K in a clathrate-like CaH$_6$ phase~\cite{Wang:2012} and the measurement of superconductivity at 203~K in H$_3$S~\cite{Drozdov:2015a}, the landscape of superconducting materials with high hydrogen content has been heavily explored -- for binary phases, primarily. The nature of the hydrogenic sublattice in these structures has been found to heavily influence their superconducting properties, with high $T_c$s belonging to phases in which H$_2$ units have been disrupted, resulting in elongated H-H distances and/or multicentered bonds between hydrogen and another element. Exemplars include phases with clathrate-like three-dimensional networks of loosely bound hydrogen and the family of MH$_4$ phases that mix atomic and H$_2$ units. In the latter, $T_c$ is further correlated with H-H distance in the molecular units, which increases with larger metal atoms, and increased electron transfer from those metal atoms that results in populating the antibonding states of the H network. Anomalously high $T_c$ values also crop up in phases with more unique H motifs, including ``pentagraphene-like'' units in ScH$_{10}$. The large values for the electron-phonon coupling constants that result in high $T_c$s are associated with an electronic structure easily perturbed by atomic vibrations as well as a large amount of H-based states near $E_F$~\cite{Tanaka:2017a,Zhang:2017}. With the vibrations of the hydrogen lattice generally spread across a wide range of frequencies, this effect is strongest for the phases with clathrate-like lattices and weaker for phases based on H$_2^{\delta-}$ units, such as the ThCr$_2$Si$_2$ MH$_4$ family.

Ternary hydrides -- or quaternary and higher -- represent a new frontier where the search is on for superconducting materials that maintain dynamic stability to lower and lower pressures~\cite{diCataldo:2021b,Zhang:2021,Liang:2021}. Thus far, investigation of ternary phase diagrams remains somewhat piecemeal, with only a few full convex hulls calculated~\cite{Shao:2019,Li:2020,diCataldo:2020,Hutcheon:2020,Zhang:2021,diCataldo:2021,Semenok:2021b}. The combinatorial possibilities for three elements generate a massive phase space that cannot be thoroughly investigated -- whether by current experimental or theoretical techniques -- on a reasonable timescale. The goal for the theoretician is thus threefold: to identify promising systems for more detailed studies~\cite{Hutcheon:2020}, as well as to develop new methods and protocols for rapid calculations, and the analysis of the large amount of data generated during the construction of full ternary convex hulls. Moreover, it is expected that entropic effects, currently neglected, will become key in the stabilization of multicomponent systems~\cite{Toher:2019} 

Towards the former goal, a common strategy is to consider substitutions of a third element into a known binary hydride, exemplified by the mixed La/Y hydrides recently shown to form~\cite{Semenok:2021b} and the MXH$_8$ phases from intercalation of a third element into a common MH$_8$ skeleton~\cite{diCataldo:2021,Zhang:2021,Liang:2021}. Systematic studies to identify trends in binary hydrides can identify plausible third ``alloying'' elements or locations in a binary lattice susceptible to substitution~\cite{Shipley:2021}. After selecting promising systems, strategies to more rapidly process the computations required for crystal structure prediction include using interatomic force fields for local optimizations rather than full DFT\cite{Hong:2020,Deringer:2018,Deringer:2018a}. Machine learning techniques can expedite the generation of useful force fields and in some cases have already been integrated into CSP packages~\cite{Tong:2018,Podryabinkin:2019}. The feedback loop of experimental and theoretical investigation into superconducting hydrides, then, may explore strange new ternary systems, seek out new structural motifs, and boldly discover superconducting phases -- at pressures and temperatures where none have been found before.

\section*{Acknowledgments}
We acknowledge the NSF (DMR-1827815) for financial support. K.\ P.\ H.\ thanks the US Department of Energy, National Nuclear Security Administration, through the Chicago-DOE Alliance Center under Cooperative Agreement Grant No.\ DE-NA0003975 for financial support.

\bibliography{Review}

\begin{thebibliography}{178}
\bibitem [{not()}]{note}%
  \BibitemOpen
  \href@noop {} {}\bibinfo {note} {In a footnote Wigner and Huntington say that
  ``It was J.\ D.\ Bernal who first put forward the view that all substances go
  over under very high presure into metallic or valence
  lattices.''}\BibitemShut {NoStop}%
\end{thebibliography}

\end{document}